\documentclass[prb,twocolumn,aps]{revtex4-2}
\usepackage{color}
\usepackage{xcolor}
\usepackage{graphicx}
\usepackage{amsmath}
\usepackage{natbib}
\usepackage{graphicx}
\usepackage{mathtools}
\usepackage{mathrsfs}
\usepackage{indentfirst}
\newcommand{\wmk}{~W~m$^{-1}$~K$^{-1}$~}

\begin{document}

\title{Mode Localization and Suppressed Heat Transport in Amorphous Alloys}
\author{Nicholas W. Lundgren}
\affiliation{Department of Chemistry, University of California, Davis, Davis, CA 95616, USA}

\author{Giuseppe Barbalinardo}
\affiliation{Department of Chemistry, University of California, Davis, Davis, CA 95616, USA}

\author{Davide Donadio}
\affiliation{Department of Chemistry, University of California, Davis, Davis, CA 95616, USA}
\email{ddonadio@ucdavis.edu}
\date{\today}
\begin{abstract} 
Glasses usually represent the lower limit for the thermal conductivity of solids, but a fundamental understanding of lattice heat transport in amorphous materials can provide design rules to beat such a limit. 
Here we investigate the role of mass disorder in glasses by studying amorphous silicon-germanium alloy (a-Si$_{1-x}$Ge$_x$) over the full range of atomic concentration from $x=0$ to $x=1$, using molecular dynamics and the quasi-harmonic Green-Kubo lattice dynamics formalism. 
We find that the thermal conductivity of a-Si$_{1-x}$Ge$_x$ as a function of $x$ exhibits a smoother U-shape than in crystalline mass-disordered alloys.
The main contribution to the initial drop of thermal conductivity at low Ge concentration stems from the localization of otherwise extended modes that make up the lowest 8\% of the population by frequency. Contributions from intermediate frequency modes are decreased more gradually with increasing Ge to reach a broad minimum thermal conductivity between concentrations of Ge from $x=0.25$ to $0.75$.
Modal analysis unravels the correlations among localization, line broadening, and the contribution to thermal transport of modes within different frequency ranges.

\end{abstract}
\maketitle

\section{Introduction}
Understanding the mechanism of heat transport in glasses at the atomistic level is essential to designing functional amorphous materials for a wide range of applications in which thermal management is a key issue, including energy conversion and storage, thermal insulation in electronics, and thermal barrier coating in turbines. 
In solids, thermal energy is carried primarily by quantized atomic vibrations, which, in crystals, can be expressed as Bloch functions and are formally referred to as phonons. 
Due to the lack of periodic order, in amorphous materials, the strict definition of phonons does not apply.
Still, quantized vibrational modes are responsible for heat transport in glasses, however through different, less efficient mechanisms than in crystalline solids.
As a consequence of disorder, some vibrational modes exhibit a certain degree of localization, which further hinders heat transport.
Because of this, for a given chemical composition, glasses usually exhibit among the lowest limit of thermal conductivity, which makes them ideal materials for thermal insulation, but also potential candidates for efficient thermoelectric energy conversion.\cite{nolas2002, mizuno2015beating}

The interpretation of heat transport in glasses is complicated by the nature of vibrational modes in disordered systems, for which standard phonon properties, such as group velocity and mean free path, cannot be defined. 
Nevertheless, atomistic simulations have identified some of the main features that control the lattice thermal conductivity of glasses.\cite{Zhou:2019fv} 
On the one hand, molecular dynamics (MD) is the preferred method to compute thermal conductivity in disordered systems, as it does not rely on the definition of phonons and entails full anharmonicity.\cite{Jund1999,Shenogin2009,sosso2012} 
On the other hand, the pioneering work by Allen and Feldman established a framework to use lattice dynamics for this purpose, which provided insight into how vibrational modes with different frequency and localization character contribute to the overall thermal conductivity of glasses.\cite{Allen:1989ua,Allen:1993uc,Feldman:1993tn,Feldman:1999uw,allen1999} 
The Allen-Feldman framework was eventually combined with the Peierls-Boltzmann theory of thermal transport in crystals to model heat transport in partially disordered materials and nanostructures.\cite{Donadio:2010kp,Zhu:2016bi,Zhu:2016ks,Zhu:2018kf,Zhou:2020jp} This practical approach successfully identified the salient features of heat transport in amorphous silicon, highlighting the combined contribution of low-frequency propagating phonons and intermediate-frequency diffusive modes to the total thermal conductivity,\cite{he2011,Larkin:2014dc,lv2016direct} though the strength of the relative contributions is still debated.\cite{Shenogin2009,braun2016size,moon2018}
The two transport models were finally unified in a general formula of thermal transport that describes equally well both ordered and disordered materials.\cite{simoncelli2019,isaeva2019} 
In particular, the quasi-harmonic Green-Kubo (QHGK) approach proved reliable to compute the thermal conductivity of crystalline, amorphous, and nanostructured silicon, showing excellent agreement with molecular dynamics simulations and available experimental measurements.\cite{isaeva2019,Neogi:2020vz,Zink:2006wt}
The advantage of QHGK over previous lattice dynamics methods is that it does not require any {\it a priori} assumption on the nature of the heat carriers, whether ``{\sl propagons}", ``{\sl diffusons}" or ``{\sl locons}",\cite{allen1999} as all the vibrational modes are treated on equal footing. This approach may then be used to carry out further unbiased studies of the physics of heat transport in glasses, considering more complex systems, for example, the effect of extrinsic defects in glasses. 
Attaining a deeper understanding of these systems would allow one to engineer glasses with designed thermal transport properties.

Alloying is an effective approach to reduce the thermal conductivity of crystalline silicon (c-Si) and nanostructures.\cite{Abeles:1963,Garg:2011hi,Wang:2010hw,larkin2013predicting,Aksamija:2013hd,Upadhyaya:2015ir,PhysRevApplied.6.014015,Xiong:2017if,FerrandoVillalba:2020hc}
The thermal conductivity $\kappa$ of c-Si as a function of Ge concentration traces a U-shape curve, with a drop in  conductivity by up to a factor of 20 ($\kappa=150$ \wmk for c-Si, $\kappa\sim 7$ \wmk for c-Si$_{1-x}$Ge$_x$ with $0.2<x<0.8$)
Such reduction stems from mass-disorder phonon scattering, which abates phonon mean free path, with especially high scattering rates for phonons with frequency higher than $\sim$2 THz.\cite{Garg:2011hi,PhysRevApplied.6.014015} 
A recent study found that Anderson localization of phonons may occur in binary isotopic alloys,\cite{Mondal:2017du} and when the light element (in our case Si) prevails, low-frequency modes are more heavily localized, and {\it vice versa}.
Equal concentrations of $0.5$ would produce the maximum localization over the spectrum. Furthermore, an increased difference between the masses causes stronger localization effects,\cite{Mondal:2017du} and lowers lattic thermal conductivity.\cite{PhysRevApplied.6.014015} 
Giri {\sl et al.} addressed heat transport in a-SiGe by a combination of equilibrium and nonequilibrium MD.\cite{giri2018localization} They interpreted the observed thermal conductivity reduction (up to $\sim 53$\%) as a combination of reduced mean free path of phonon-like modes and lowered diffusivity of {\sl diffusons}, according to the calssification proposed in Ref.~\onlinecite{allen1999}. 
Mode localization was proposed as the mechanism that hinders heat transport in glassy Si/Ge superlattices, but its relation to the reduction of $\kappa$ upon alloying was not clarified. 

In this article, we investigate the effect of alloying and mass-disorder on lattice thermal transport in amorphous silicon.
We use molecular dynamics (MD) and QHGK lattice dynamics to compute the thermal conductivity of a set of different models, obtained by simulating quenching from the melt and random Ge substitutions with the full range of Ge concentrations from 0 to 1.
The thermal conductivity is computed by both QHGK and MD, and we analyze the contribution of the vibrational modes to the total thermal conductivity. 
Modal analysis, enabled by QHGK, shows that the thermal conductivity reduction is tightly connected to phonon localization induced by mass disorder, which affects to the largest extent the lowest-frequency part of the vibrational spectrum of a-Si$_{1-x}$Ge$_x$.

\section{Methods}
In the MD simulations and lattice dynamics calculations, we model amorphous Si$_{1-x}$Ge$_x$ using the many-body Tersoff potential.\cite{tersoff1989modeling}
Silicon glass models were generated through MD simulations of quenching from the melt.
MD trajectories are computed using the velocity Verlet algorithm as implemented in LAMMPS\cite{plimpton1993fast} with a timestep of 1~fs. For simulations in the constant-volume canonical ensemble (NVT) the temperature is controlled by stochastic velocity rescaling,\cite{bussi2007canonical} whereas for the constant pressure canonical ensemble (NPT) the Nose-Hoover algorithm is employed.\cite{Martyna:1992gy} 
Crystalline silicon models were molten at 4000~K for 3~ns, and then quenched to 2000~K, about 20\% below the freezing point of the model, over 5~ns in line with the work of Ishimaru et. al.\cite{ishimaru1997generation}. 
The models were then annealed for 30~ns, while being cooled at a constant rate from 2000~K to 1400~K, and were finally brought to 300~K over 5~ns. All these steps were carried out in the NVT ensemble at the fixed density of 2.32~g/cm$^3$
Lastly, the models were run for 1~ns at constant pressure. In total, four models of 1728 atoms and single models containing 4096 and 13824 atoms were produced for a total of six glass structures.

Ge doped alloy models were obtained by replacing randomly selected silicon atoms with germanium atoms.
The structures were then optimized at zero temperature \textcolor{black}{and at constant volume} using a quasi-Newton optimization algorithm\cite{sheppard2008optimization} imposing the Frobenius norm of interatomic forces below 1e-6~eV/\AA.
Each model has ten variants with Ge concentrations ranging from 0 to 100\% with a finer sampling for configurations nearer to pure a-Si.

We first computed the room temperature thermal conductivity $\kappa$ of the 1728 atom systems by equilibrium MD, evaluating the Green-Kubo integral of the heat flux $\hat J(t)$ autocorrelation function:\cite{ZWANZIG:1965tp}
\begin{equation}
    \kappa = \frac{1}{K_B T^2 V} \int^{\infty }_0 \left< J (t),  J(0)\right> dt,
    \label{eq:gk}
\end{equation} 
where $k_B$ is the Boltzmann constant, $T$ is the temperature and $V$ is the volume. 
$\kappa$ is evaluated using cepstral analysis, which eases the time-convergence issue of taking the infinite time integral in Equation \ref{eq:gk} by estimating the zero-frequency limit of the log-spectrum of the heat flux computed over a finite time MD simulation.\cite{Ercole:2017eea}
The cepstral analysis, performed with the thermocepstrum code, was carried out on heat flux outputs of 10~ns microcanonical equilibrium MD trajectories integrated with a 0.5~fs timestep using the GPUMD package,\cite{fan2017efficient} which implements the correct partitioning of the many-body forces to compute $\hat J(t)$.\cite{Fan:2015ba}
The sampling frequency for calculating the power spectral density in the cepstral analysis was 12~THz.

The QHGK calculations for the thermal conductivity of a-SiGe systems were performed with the open-source lattice dynamics calculator $\kappa$ALDo.\cite{barbalinardo2020efficient}
The models were studied at 300~K with quantum mechanical treatment of heat capacities and phonon populations.
In QHGK the thermal conductivity is computed as,
\begin{equation}
    \kappa_{\alpha \beta} = \frac{1}{V} \sum_{nm} c_{nm} v_{nm}^{\alpha} \tau^{\circ}_{nm} v_{nm}^{\beta},
\label{eq:qhgk}
\end{equation}
where we introduced the generalized specific heat $c_{nm}$, the generalized lifetime $\tau^{\circ}_{nm}$, and the generalized velocity $v_{mn}^\alpha$.
\textcolor{black}{
The subscripts $n$ and $m$ refer to the indexes of the eigenvectors $\mathbf{e_n}$ and $\mathbf{e_m}$ of the dynamical matrix $\mathcal{D}$ of the system, and $v_{mn}^\alpha$ :
\begin{equation}
    v^\alpha_{nm} = \frac{1}{2 \sqrt{\omega_{n} \omega_{m} }} \langle \mathbf{e}_n |   
    \mathcal{S}^\alpha |\mathbf{e}_n \rangle
\label{eq:vnm}
\end{equation}
is proportional to the matrix element of the operator representing the heat flux in the direction $\alpha$  in the harmonic approximation $S^\alpha_{i\delta,j\gamma}=\left( R^{\circ}_{i\alpha} - R^{\circ}_{j\alpha}\right) \mathcal{D}_{i\delta}^{j\gamma} $.\cite{Allen:1989ua,isaeva2019} 
}
The generalized lifetimes are defined by
\begin{equation}
    \tau _{nm}^ \circ = \frac{{\gamma _n + \gamma _m}}{{(\gamma _n + \gamma _m)^2 + (\omega _n - \omega _m)^2}} + {\cal{O}}(\epsilon ^2),
\label{eq:tau}
\end{equation}
where $\gamma_{n}$ is the decay rate (or linewidth), and $\omega_{n}$ is the frequency of mode $n$, and $\epsilon$ is the ratio $\gamma / \omega$. We stress that the generalized lifetimes naturally account for the mass difference scattering without the need for {\it ad hoc} scattering terms or empirical parameters. 
Phonon linewidths are calculated explicitly for the 1728 and 4096 atom systems using Fermi's Golden rule.\cite{Fabian:2003fv} The linewidths of the 13824 atom system were interpolated from a third-order spline curve fit of the 4096 atom system. The details of this procedure have been described in a previous work.\cite{isaeva2019}
Anharmonic lattice dynamics calculations are carried out computing the second and third derivatives of the potential for the optimized models by finite differences with atomic displacements of 1e-5~\AA.
Because the Tersoff potential is smoothed to zero by a sigmoid function between 2.7 and 3.1 \AA, in amorphous models, atoms falling in this transition region may be subject to an unphysically steep potential that introduces large numerical errors in the calculation of second and third derivatives. For this reason, in lattice dynamics calculations, we cut off the potential at 2.7 \AA.
QHGK calculations using the Tersoff potential provide  excellent agreement with experimental measurements of thermal conductivity at temperatures below 600 K, without the need of adjustable parameters.\cite{isaeva2019,Zink:2006wt,Cahill:1994}

Using the QHGK approach, we can also calculate the {\sl diffusivity} with the last three terms of Eq.~\ref{eq:qhgk}, which quantifies the rate of heat transfer in units of area per time for each mode. It is defined as
\begin{equation}
    D_{n} = \frac{1}{N_{atoms} V} \sum_{m} v_{nm}^{\alpha} \tau^{\circ}_{nm} v_{nm}^{\beta},
\label{eq:diff}
\end{equation}
where $N_{atoms}$ is the number of atoms. 
The localization of the vibrational modes was evaluated by calculating the participation ratio (PR)
\begin{equation}
     \mathrm p (n) = \left(N_{atoms} \sum^{N_{atoms} } _{m=1} \vert e_m(n) \vert ^4 \right)^{-1},
    \label{eq:partratio}
\end{equation}
where $e_m(n)$ is a norm of the vector comprised of the three Cartesian components of the m$^{th}$ eigenvector acting on atom n.\cite{schober1996low}

\section{Results and discussion}
\subsection{a-Si$_{1-x}$Ge$_x$ Structures}

\begin{figure}
\begin{center}
\includegraphics[width=1.0\linewidth]{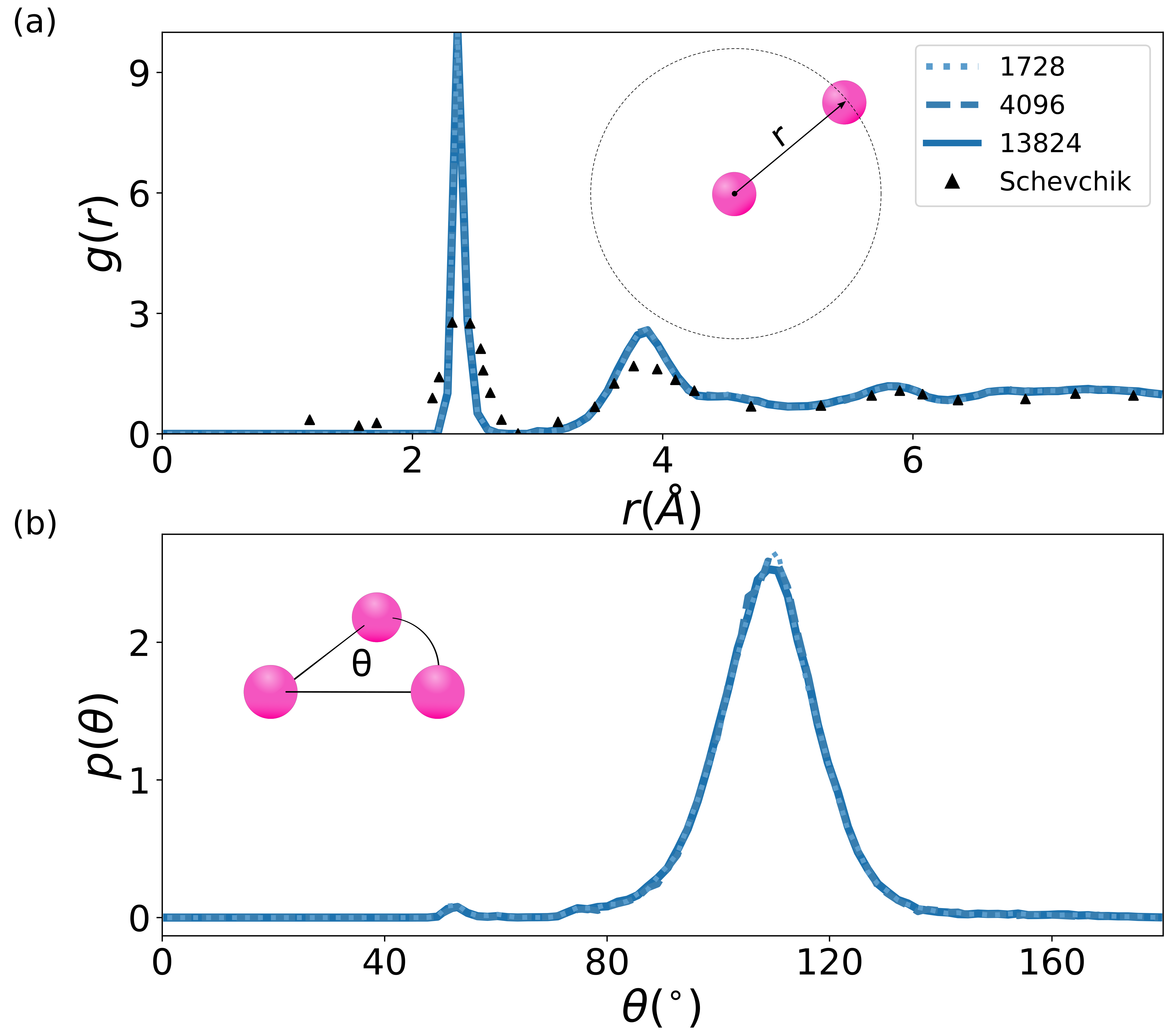}
\end{center}
\linespread{1}
\caption{
 Structural features of the  1728, 4096 and 13824 amorphous silicon: (a) radial distribution function, compared to that measured for a-SiGe by X-ray scattering;\cite{shevchik1973} (b) Angular distribution function.}
\label{fig:struc}
\end{figure}

In Figure~\ref{fig:struc} we provide the atomistic structural features of 1728, 4096 and 13824 atom amorphous silicon models, in terms of radial distribution function $g(r)$ (RDF) and angular distribution function (ADF). 
The structure of this model is representative of all the other models considered in this work. The glass structure consists of a tetrahedral random network, with minimal occurrence of 5-fold over-coordinated sites. The RDF agrees well with those measured experimentally for a-Si, a-Ge and a-SiGe glasses.\cite{kugler1989, shevchik1973, graczyk1973scanning}
In particular, Figure~\ref{fig:struc}a shows that the peak positions of the RDF of our model correspond to those measured by x-ray scattering for a-SiGe, and the two RDFs feature the same range of ordered shells, up to $\sim 6.5$~\AA. However, the experimental peaks have lower intensities than the simulations. In particular the measured nearest neighbor peak, centered at 2.3~\AA\ is much lower and slightly broader, but these differences may stem from the lack of correction for termination effects in the X-ray scattering experiments.\cite{shevchik1973,kazimirov2000termination}
It is important to note that experiments show that there are no substantial structural differences in the structure of these glasses, which justifies the procedure, through which we generate amorphous alloys by random substitution of Si atoms with Ge.  

Experimental techniques to measure angular distribution functions of multicomponent amorphous materials are still being developed, though former works suggested that the root mean square deviation from the tetrahedral angle of the angular distribution determined by Raman scattering lies between 11 and 15 degrees:\cite{maley1988dynamics, vink2001raman} 
the  12.0 degree width of the distribution in Figure~\ref{fig:struc}b falls in this experimental range. 
Finally, we stress that neither RDF nor ADF exhibit significant size effects, as they are indistinguishable for all the models considered.

\subsection{Thermal Conductivity}

\begin{figure}
\begin{center}
\includegraphics[width=1.0\linewidth]{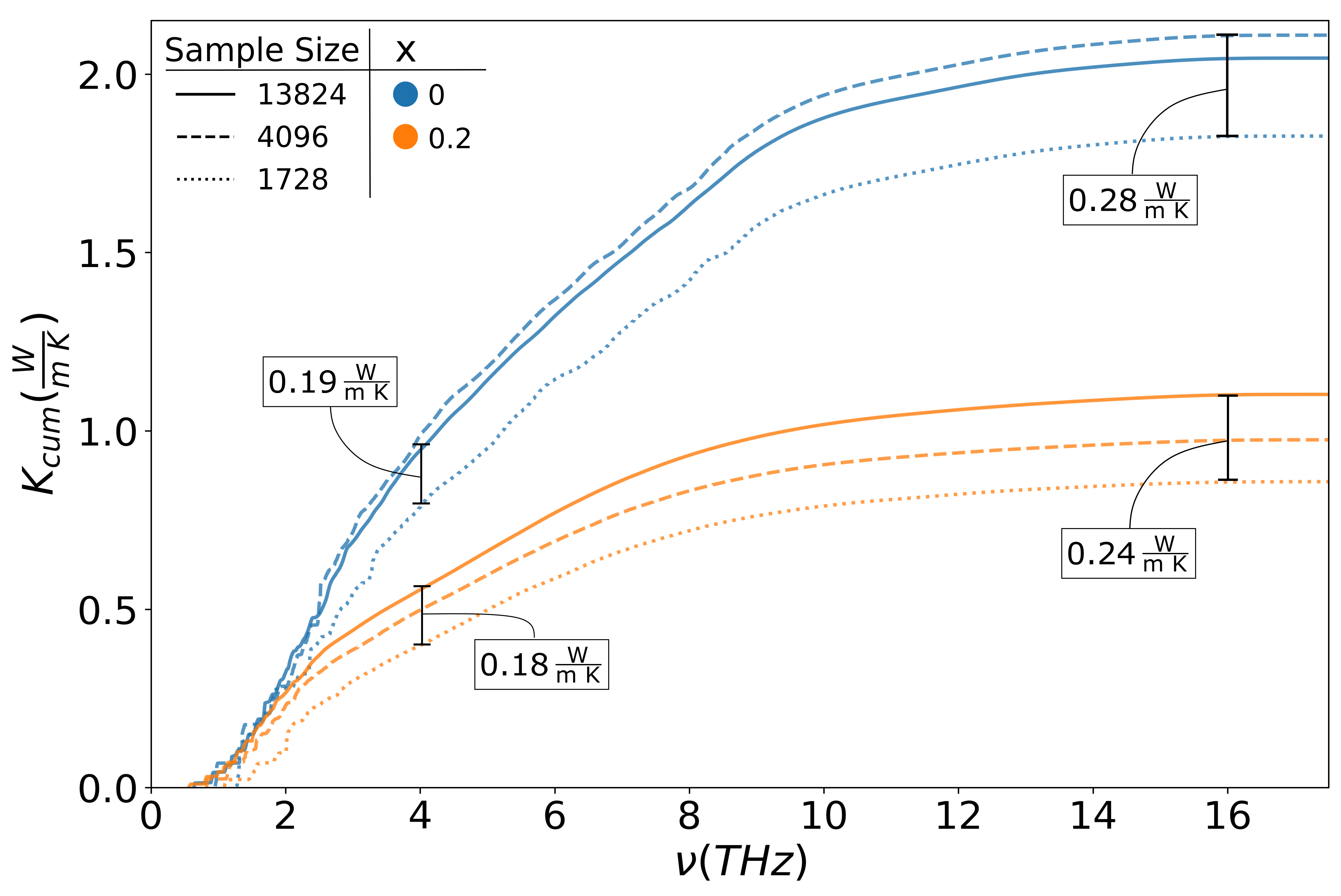}
\end{center}
\caption{
 Cumulative thermal conductivity of a-Si and a-Si$_{0.8}$Ge$_{0.2}$ models containing 1728 (dotted line), 4096 (dashed line) and 13824 atoms (solid line) as a function of increasing frequency for a-Si (blue lines) and for a-Si$_{0.8}$Ge$_{0.2}$ (orange lines).  
 }
\label{fig:size}
\end{figure}
Size convergence needs to be carefully checked in the calculations of $\kappa$ either by EMD or by lattice dynamics methods, such as QHGK.\cite{sellan2010size}
To this aim, we have computed the thermal conductivity of models containing 1728, 4096 and 13824 atom using QHGK. The comparison of the cumulative $\kappa$ as a function of frequency (Fig. \ref{fig:size}) for these three system sizes for a-Si and a-Si$_{0.8}$Ge$_{0.2}$ shows that size effects are similar irrespective of the composition. In particular, the differences between the small systems of 1728 atoms and the well-converged 13824 atom systems mostly build in the low-frequency part of the spectrum, below 4 THz, and are roughly the same for both compositions. This means that we can investigate thermal conductivity trends at room temperature using relatively small 1728-atom models, while accurate quantitative estimates of $\kappa$ and of the modal contribution, especially from low-frequency modes, need to be carried out for larger models.

\begin{figure}[b]
\begin{center}
\includegraphics[width=1.\linewidth]{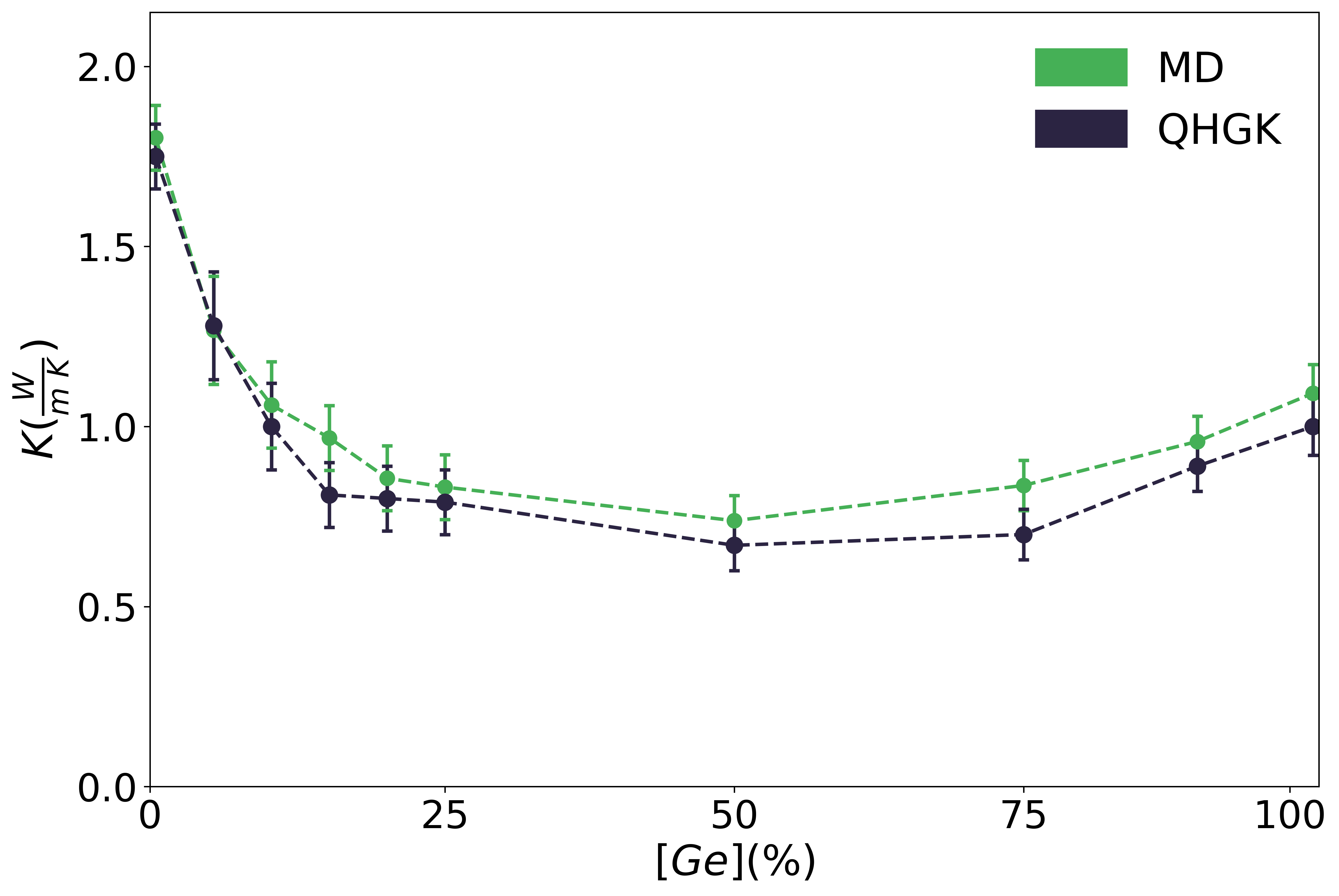}
\end{center}
\caption{Room temperature thermal conductivity of a-Si$_{1-x}$Ge$_{x}$ as a function of Ge concentration $x$\%, computed by anharmonic lattice dynamics QHGK (black) and equilibrium MD (green). 
The MD error bars are computed as the average error obtained by {\it cepstral analysis}\cite{Ercole:2017eea} of 10~ns trajectories for each of the four models. The QHGK error bars represent the standard deviations of single conductivity calculation for the same four models.}
\label{fig:k}
\end{figure}
Figure~\ref{fig:k} displays the thermal conductivity as a function of the Ge concentration $x$, calculated using either equilibrium MD or the QHGK formula with classical phonon populations. The two results agree very well within the confidence interval obtained by the variance over ten different models and the statistical error in the evaluation of $\kappa$ from the cepstral analysis of the heat current time series from equilibrium MD. Such an agreement allows us to use QHGK with confidence to compute and analyze thermal transport in larger a-Si$_{1-x}$Ge$_x$ models using the correct quantum statistics for phonons. 

\begin{figure}[bt]
\begin{center}
\includegraphics[width=1.\linewidth]{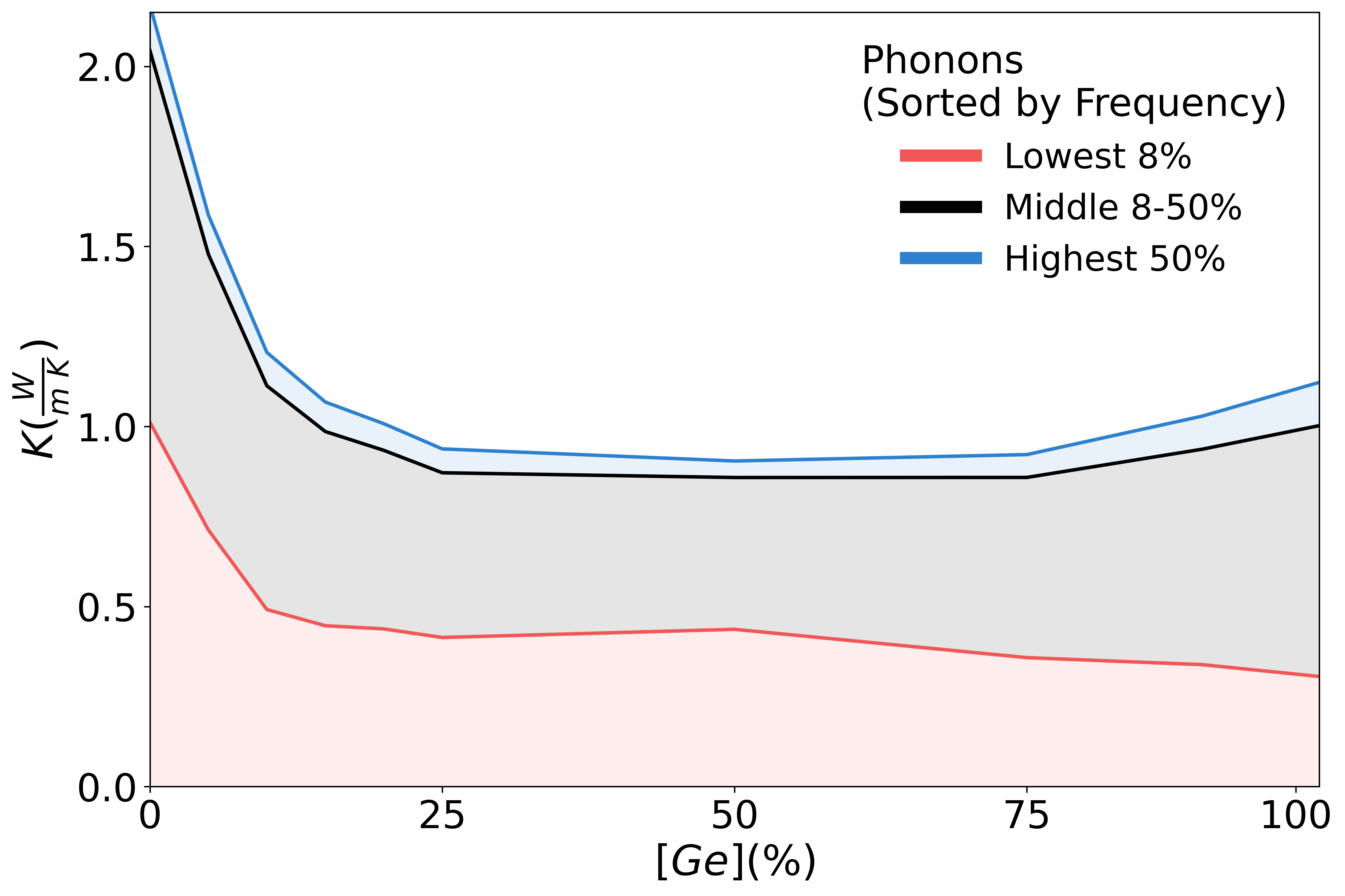}
\end{center}
\caption{
    Thermal conductivity of the 4096 atom, a-Si$_{1-x}$Ge$_x$ alloy models as a function of Ge concentration, split into contribution by frequency ranges. The red, black, and blue regions represent the contributions from the 8\% lowest frequency modes, the 8--50\% modes with intermediate frequency, and the remaining 50\% highest frequency modes.
    These frequency ranges correspond to $<4$ THz, between 4 and 10 THz and $>10$ THz for a-Si, and they shift toward lower frequencies as the vibrational density of states shifts when Ge is increased (see Figure~\ref{fig:obs}b). 
}
\label{fig:freq_breakdown}
\end{figure}

Calculations on a 4096 atom model (Figure~\ref{fig:freq_breakdown}) show that $\kappa(x)$ exhibits a U-shape pattern, qualitatively similar, but much shallower than that found for the crystal analogue.\cite{Abeles:1963}  
In agreement with previous calculations,\cite{giri2018localization} the maximum thermal conductivity reduction produced by alloying a-Si with Ge is about a factor 2. 
This is relatively small compared to the case of c-SiGe  mentioned for which $\kappa$ drops by more than 20 times.\cite{Abeles:1963,Garg:2011hi}
According to our simulations, starting from a reference $\kappa=2.2$~\wmk for a-Si, the substitution of 5\% Ge results in a 0.6~\wmk drop of $\kappa$. A further 0.55~\wmk reduction of $\kappa$ is obtained raising the Ge concentration to 15\%, but increasing germanium content beyond 20\% does not produce significant changes in the thermal conductivity. 
Similar to the cases of crystalline and nanostructured silicon,\cite{Garg:2011hi, Wang:2010hw, He:2011ig, Xiong:2017if} this trend indicates that the effect of alloying on lattice thermal transport is prominent at low Ge concentrations and tends to saturate for $x>0.1$. However, in a-SiGe we cannot attribute the thermal conductivity change to phonon scattering from mass disorder, as the heat transport mechanism in glasses is different from crystalline materials.

Hereafter we exploit the detailed information obtained from lattice dynamics analysis and the QHGK formalism to resolve the contribution of each vibrational mode to the total thermal conductivity of a-Si$_{1-x}$Ge$_x$ as a function of $x$. 
Figure~\ref{fig:freq_breakdown} shows the cumulative thermal conductivity of a-Si$_{1-x}$Ge$_x$ as a function of the concentration $x$, split into lower, middle and higher frequency contributions. The three frequency ranges are defined in terms of the number of modes as the density of vibrational states changes as a function of the Ge content, with the low-frequency range consisting of 8\% of the modes, the middle range of 8--50\% of the modes and the high-frequency range the remnant top 50\%.  
In a-Si the modes with frequencies less than 10~THz contribute to about 95\% of the total $\kappa$, despite representing only 50\% of the population. Of these, the lowest 8\% of the population, comprising the 0-4~THz region, contribute about half of the total conductivity, in agreement with previous work that identified this frequency range as the most important to thermal conduction.\cite{he2011,Larkin:2014dc,isaeva2019} 
\textcolor{black}{The other half of $\kappa$ is supplied by the mid-range 8--50\% of the modes.}
The last 50\% of the population contributes only $\sim$5\% of the total conductivity, making them negligible in comparison.
$x=0.05$ Ge substitution reduces the thermal conductivity contribution of the 0--4 THz modes by $\sim$30\%, accompanied by a $\sim$25\% reduction in the contribution from the mid-range frequencies. Yet, the latter group contains many more vibrational modes than the former, providing a higher contribution in total.
The somewhat negligible contributions of the high-frequency modes do not change significantly as $x$ is increased.
\textcolor{black}{Increasing $x$ to 0.25 produces an overall $\sim 50$\% reduction of the contribution to $\kappa$ from both low- and mid-frequency modes, leading to the minimum $\kappa$ plateau that extends up to x=0.75.}
\textcolor{black}{Remarkably, the contribution from low-frequency modes is not restored by adding Ge beyond $x=0.5$, and it even keeps decreasing slightly to reach a minimum for pure amorphous Ge. In turn, the contribution from mid-frequency modes increases and it becomes prominent thus making the $\kappa (x)$ curve asymmetric.
The lower thermal conductivity of a-Ge with respect to a-Si stems from the larger mass of Ge, and, to a lesser extent, from the larger anharmonicity of the Ge-Ge bonds. Larger mass leads to a reduction of the generalized group velocities and the increased anharmonicity abates lifetimes. These effects are more prominent on the low-frequency modes, which have a propagating character.}

\begin{figure}[b]
\begin{center}
\includegraphics[width=1.\linewidth]{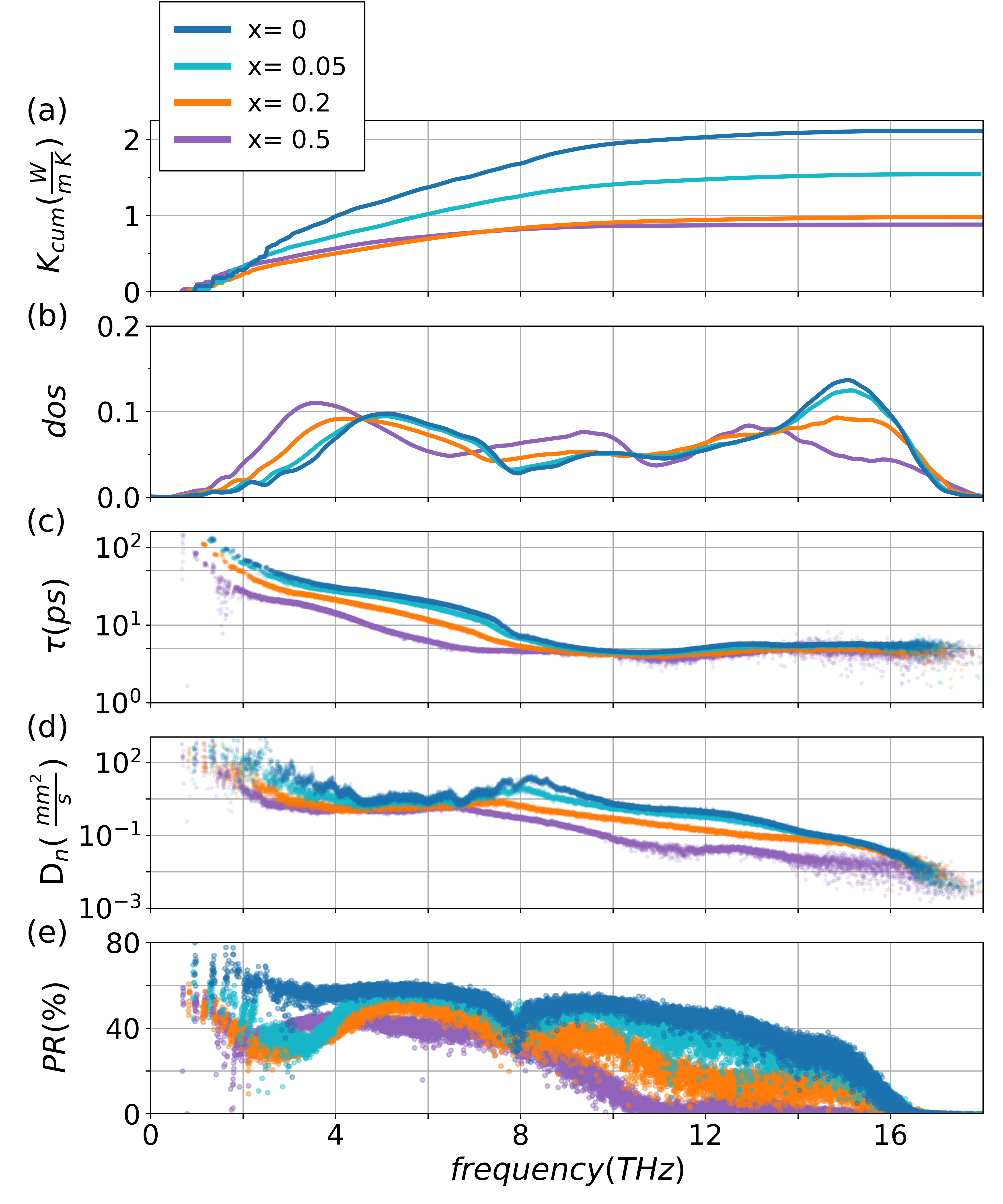}
\end{center}
\caption{
Modal analysis of the 4096-atom a-Si$_{1-x}$Ge$_x$ model for $x=0$, 5, 20 and 50: a) cumulative conductivity, b) density of states, c) lifetimes, d) modal diffusivity, and e) participation ratio for each normal mode plotted against frequency.
}
\label{fig:obs}
\end{figure}

\begin{figure}
\begin{center}
\includegraphics[width=1.\linewidth]{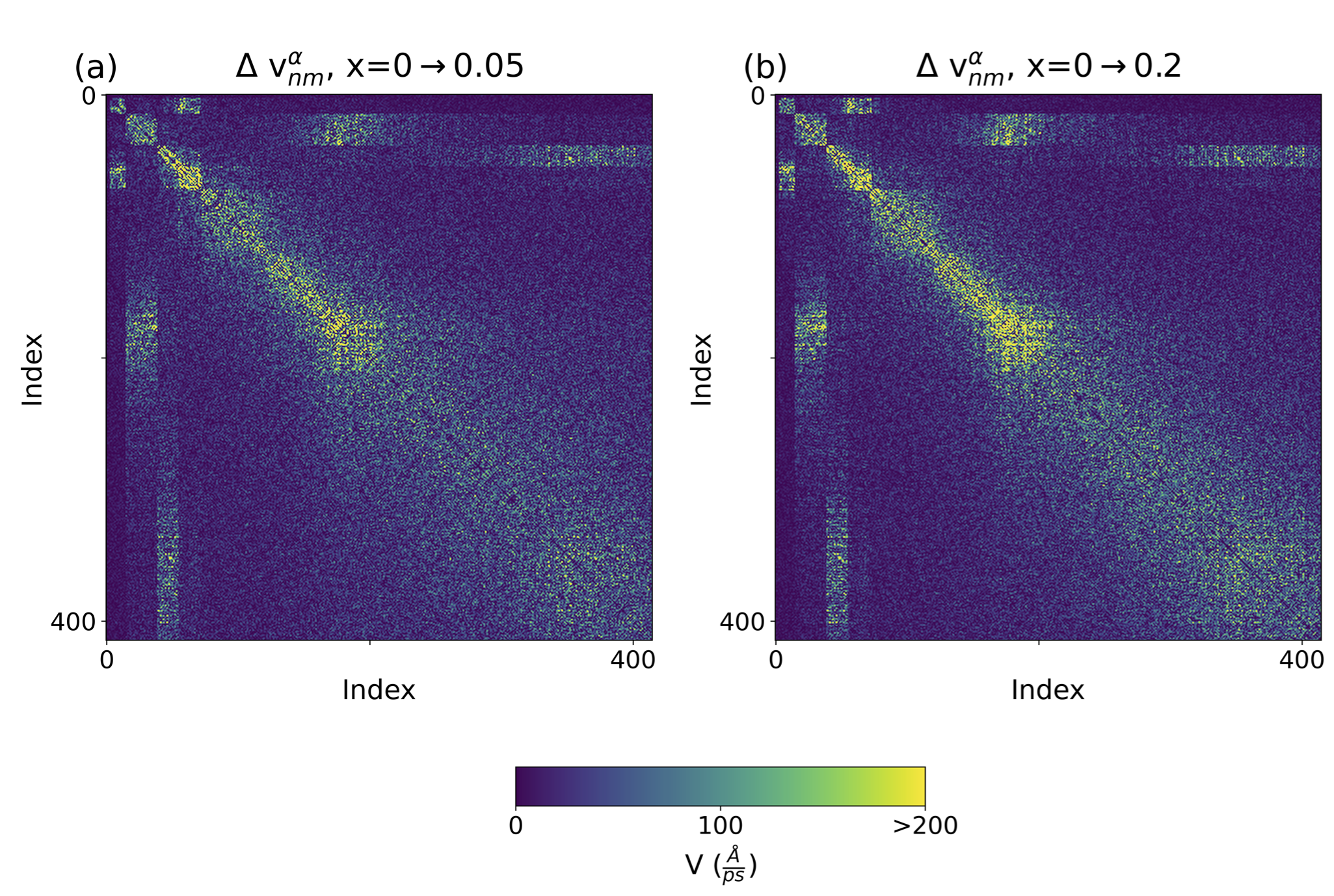}
\end{center}
\caption{\textcolor{black}{Difference in the norm of the generalized velocities between a-Si and a-Si$_{1-x}$Ge$_x$ with $x=0.05$ and $x=0.2$ in the low frequency block diagonal.}
}
\label{fig:vnm}
\end{figure}

The spectral contribution to heat transport is further analyzed in Figure~\ref{fig:obs}a, in which the cumulative thermal conductivity for $x=$0, 0.05, 0.2 and 0.5 is plotted against frequency. Thermal conductivity differences among the four models are more dominant at low frequency and remain roughly constant for frequencies larger than 8~THz. There is only a minimal difference between the models with Ge concentration of 0.2 and 0.5. 
The phonon density of states (Fig.~\ref{fig:obs}b) shows that increasing the germanium content causes an uptick in the phonon population in the low-frequency region, \textcolor{black}{due to the larger atomic mass of Ge atoms and the slightly softer Ge-Ge and Si-Ge bonds.}
However, the increasing number of low-frequency modes comes with a decrease in their ability to transfer heat, as seen in the accumulation function. 
Figure~\ref{fig:obs}c shows the vibrational modes lifetimes $\tau$ and shows that the large reduction of $\kappa$ upon alloying a-Si with 5\% Ge is not associated with a significant reduction of $\tau$. 
Hence, the main mechanism that hinders heat transport in the alloy glass at low Ge content is not mass scattering, but it is related to changes in the generalized velocities (Eq.~\ref{eq:qhgk}), which account for resonant transport among delocalized modes. 
\textcolor{black}{Figure~\ref{fig:vnm} shows the absolute changes in the generalized velocity matrix $\Delta v_{nm}$ for the first 400 low-frequency modes (bottom 8\%) upon alloying with Ge concentration $x=0.05$ and 0.2. The $v_{nm}$ reduction is particularly significant at the lowest frequencies and it is the main mechanism that leads to the observed overall $\kappa$ reduction.}
Conversely, at higher concentrations of Ge, lifetimes are significantly reduced by mass disorder. Such reduction of $\tau$, which implies a broadening of the spectral linewidth, reduces the efficiency of the resonant heat transport mechanism. This effect combines with a reduction of generalized group velocities and leads to an overall further decrease of the low- and mid-frequency modal contributions to $\kappa$. 
\textcolor{black}{ The modal diffusivity $D_n$ (Fig.~\ref{fig:obs}d) convolutes the reduction of $v_{nm}$ with the increased line broadening. A significant decrease in $D_n$ with the increase of Ge concentration is observed both in the low and intermediate frequency range.}
%
Figure~\ref{fig:obs}e displays the modal participation ratio, which is a measure of modes' delocalization.\cite{schober1996low} The spectrum of a-Si features mostly delocalized extended modes, except for a small subset of optical modes at high-frequency.\cite{Feldman:1993tn} In particular low-frequency vibrational modes extend over the whole model and this feature accounts for their large contribution to heat transport. 
The effect of Ge alloying is the enhancement of spatial localization over the whole frequency spectrum. In particular, localization at low frequency occurs already at low Ge concentration. 

\begin{figure}
\begin{center}
\includegraphics[width=1.\linewidth]{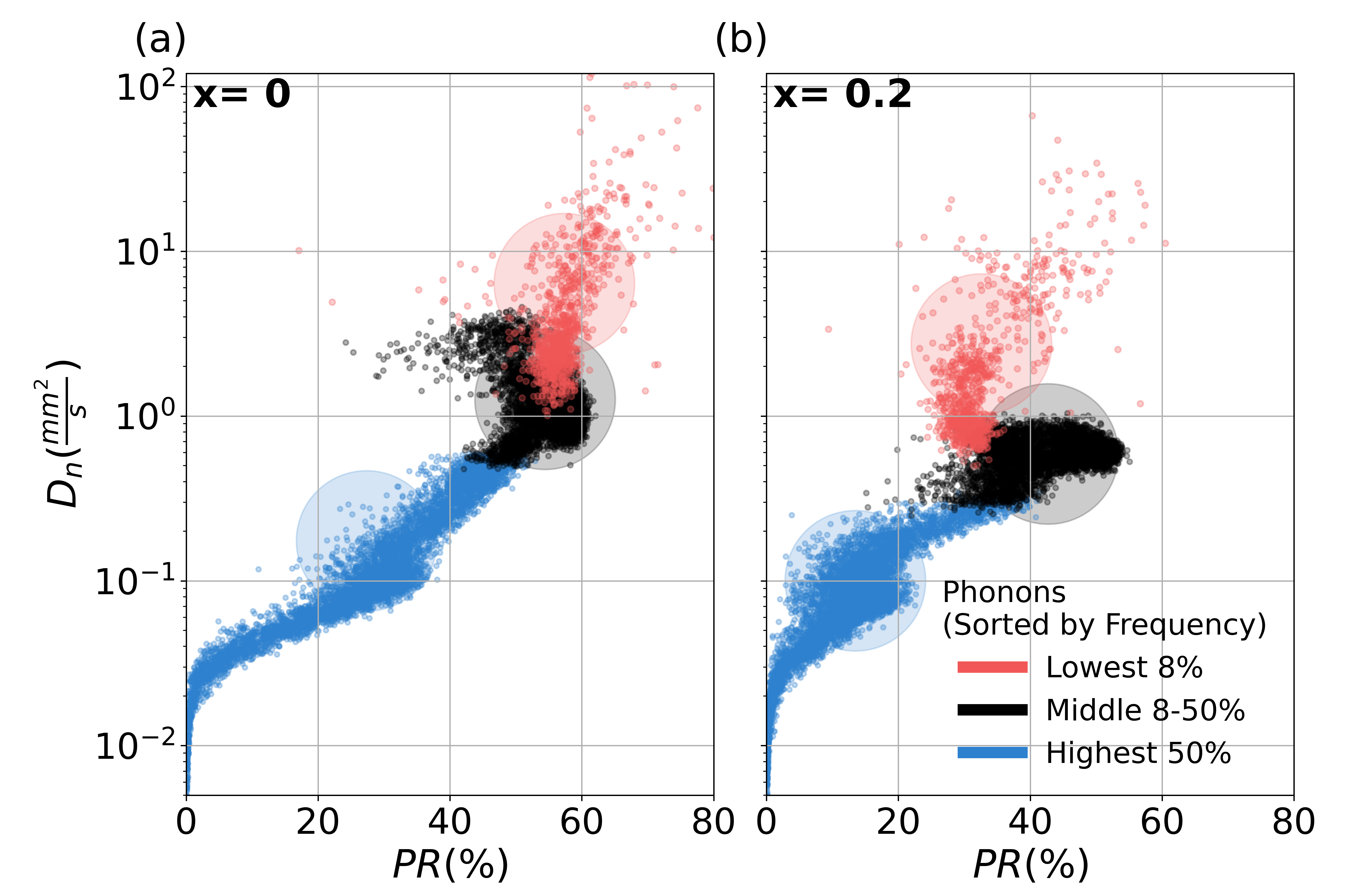}
\end{center}
\caption{
Correlation between mode diffusivity $D$ (see Eq.~\ref{eq:diff}) and mode delocalization (participation ratio, $PR$ (\%)) for the 4096 atom pure a-Si (a) and a-Si$_{0.8}$Ge$_{0.2}$ (b) models. Data are colored according to the frequency ranges defined in Figure~\ref{fig:freq_breakdown}. \textcolor{black}{Transparent bubbles indicate the center of the respective distributions.}
}
\label{fig:pr_diff}
\end{figure}
As the generalized velocities, and therefore the mode diffusivity defined in Eq.~\ref{eq:diff}, are related to the spatial overlap of vibrational modes with similar frequency, \textcolor{black}{which is necessary to make the matrix elements in Eq.~\ref{eq:vnm} finite, thus enabling  energy transport among different modes.}
It is therefore reasonable to hypothesize that mode localization is the main physical mechanism through which $\kappa$ is suppressed in glassy alloys. 
To explore the relationship between modes localization and their contribution to $\kappa$, in Figure~\ref{fig:pr_diff}, modal diffusivity is plotted against the participation ratio for pure a-Si and the 20\% Ge model, and data are color-coded according to their previous classification of low (red), mid (black) and high (blue) frequency. 
The graph shows that there is indeed a correlation between localization and diffusivity, but modes in different frequency ranges exhibit different trends, and get affected in different ways by Ge alloying. Consistent with our previous analysis, the low-frequency modes are the most affected by mass disorder, as they get localized and their diffusivity decreases accordingly. Interestingly, the diffusivity of the mid-range frequency modes is also suppressed but with less significant localization, thus suggesting that spatial localization is not the only mechanism responsible for the observed $\kappa$ reduction, but line broadening contributes as well. \textcolor{black}{For the 20\% Ge model, the contribution of these modes to $\kappa$ is nearly halved with respect to a-Si, and it is roughly equivalent to that of the low-frequency modes.} 
High-frequency modes get more localized, but their small contribution to heat transport does not vary significantly.

\section{Conclusion}
In summary, our MD and lattice dynamics simulations predict that the thermal conductivity reduction from alloying in glasses is much less prominent than in crystals, although with a similar trend of $\kappa$ as a function of concentration. 
In particular, we observe that a small concentration of heavy Ge atoms may halve the already low thermal conductivity of a-Si, but further increasing such concentration has minimal effects on $\kappa$.

In contrast with the effect of mass disorder on lattice thermal transport in crystalline alloys, which is customarily interpreted in terms of phonon-mass scattering,\cite{Garg:2011hi} modal analysis of a-Si$_{1-x}$Ge$_x$ shows that thermal conductivity reduction in glassy alloys mostly stems from the localization of low- and mid-frequency vibrational modes, accompanied by spectral line broadening. The latter reduces the efficacy of the resonance mechanism through which heat is transferred in disordered materials.  
This finding for glassy Si$_{1-x}$Ge$_x$ complements the observation of phonon localization in mass-disordered crystalline alloys\cite{Mondal:2017du} and highlights the general connection between the localization of lattice vibrations and thermal transport.

Although our numerical experiments target a specific system, we argue that the observed mode localization and suppression of thermal transport are general and intertwined phenomena that may be exploited to engineer the thermal conductivity of amorphous solids, including ceramic, polymer and macromolecular glasses, by harnessing mass disorder. 
We envisage a direct impact of these findings for several technological applications, such as the development of more efficient dense ceramics for thermal barrier coating,\cite{ClarkeTBC,clarke_oechsner_padture_2012} and the optimization of heat management in plastic commodities, organic electronics and organic thermoelectric materials.\cite{Bubnova:2012dw,Ruscher.3.125604,Cappai:2020bh}

\section{Acknowledgements}
The authors would like to acknowledge Zekun Chen for his support and suggestions. G. B. gratefully acknowledges support by the Investment Software Fellowships (grant No. OAC-1547580-479590) of the NSF Molecular Sciences Software Institute (MolSSI) (grant No. OAC-1547580) at Virginia Tech.


\begin{thebibliography}{63}%
\makeatletter
\providecommand \@ifxundefined [1]{%
 \@ifx{#1\undefined}
}%
\providecommand \@ifnum [1]{%
 \ifnum #1\expandafter \@firstoftwo
 \else \expandafter \@secondoftwo
 \fi
}%
\providecommand \@ifx [1]{%
 \ifx #1\expandafter \@firstoftwo
 \else \expandafter \@secondoftwo
 \fi
}%
\providecommand \natexlab [1]{#1}%
\providecommand \enquote  [1]{``#1''}%
\providecommand \bibnamefont  [1]{#1}%
\providecommand \bibfnamefont [1]{#1}%
\providecommand \citenamefont [1]{#1}%
\providecommand \href@noop [0]{\@secondoftwo}%
\providecommand \href [0]{\begingroup \@sanitize@url \@href}%
\providecommand \@href[1]{\@@startlink{#1}\@@href}%
\providecommand \@@href[1]{\endgroup#1\@@endlink}%
\providecommand \@sanitize@url [0]{\catcode `\\12\catcode `\$12\catcode
  `\&12\catcode `\#12\catcode `\^12\catcode `\_12\catcode `\%12\relax}%
\providecommand \@@startlink[1]{}%
\providecommand \@@endlink[0]{}%
\providecommand \url  [0]{\begingroup\@sanitize@url \@url }%
\providecommand \@url [1]{\endgroup\@href {#1}{\urlprefix }}%
\providecommand \urlprefix  [0]{URL }%
\providecommand \Eprint [0]{\href }%
\providecommand \doibase [0]{https://doi.org/}%
\providecommand \selectlanguage [0]{\@gobble}%
\providecommand \bibinfo  [0]{\@secondoftwo}%
\providecommand \bibfield  [0]{\@secondoftwo}%
\providecommand \translation [1]{[#1]}%
\providecommand \BibitemOpen [0]{}%
\providecommand \bibitemStop [0]{}%
\providecommand \bibitemNoStop [0]{.\EOS\space}%
\providecommand \EOS [0]{\spacefactor3000\relax}%
\providecommand \BibitemShut  [1]{\csname bibitem#1\endcsname}%
\let\auto@bib@innerbib\@empty
\bibitem [{\citenamefont {Nolas}\ and\ \citenamefont
  {Goldsmid}(2002)}]{nolas2002}%
  \BibitemOpen
  \bibfield  {author} {\bibinfo {author} {\bibfnamefont {G.}~\bibnamefont
  {Nolas}}\ and\ \bibinfo {author} {\bibfnamefont {H.}~\bibnamefont
  {Goldsmid}},\ }\bibfield  {title} {\bibinfo {title} {The figure of merit in
  amorphous thermoelectrics},\ }\href@noop {} {\bibfield  {journal} {\bibinfo
  {journal} {Phys. Status Solidi A}\ }\textbf {\bibinfo {volume} {194}},\
  \bibinfo {pages} {271} (\bibinfo {year} {2002})}\BibitemShut {NoStop}%
\bibitem [{\citenamefont {Mizuno}\ \emph {et~al.}(2015)\citenamefont {Mizuno},
  \citenamefont {Mossa},\ and\ \citenamefont {Barrat}}]{mizuno2015beating}%
  \BibitemOpen
  \bibfield  {author} {\bibinfo {author} {\bibfnamefont {H.}~\bibnamefont
  {Mizuno}}, \bibinfo {author} {\bibfnamefont {S.}~\bibnamefont {Mossa}},\ and\
  \bibinfo {author} {\bibfnamefont {J.-L.}\ \bibnamefont {Barrat}},\ }\bibfield
   {title} {\bibinfo {title} {Beating the amorphous limit in thermal
  conductivity by superlattices design},\ }\href@noop {} {\bibfield  {journal}
  {\bibinfo  {journal} {Sci. Rep.}\ }\textbf {\bibinfo {volume} {5}},\ \bibinfo
  {pages} {14116} (\bibinfo {year} {2015})}\BibitemShut {NoStop}%
\bibitem [{\citenamefont {Zhou}\ \emph {et~al.}(2019)\citenamefont {Zhou},
  \citenamefont {Cheng}, \citenamefont {Chen}, \citenamefont {Xie},
  \citenamefont {Wang},\ and\ \citenamefont {Zhang}}]{Zhou:2019fv}%
  \BibitemOpen
  \bibfield  {author} {\bibinfo {author} {\bibfnamefont {W.~X.}\ \bibnamefont
  {Zhou}}, \bibinfo {author} {\bibfnamefont {Y.}~\bibnamefont {Cheng}},
  \bibinfo {author} {\bibfnamefont {K.-Q.}\ \bibnamefont {Chen}}, \bibinfo
  {author} {\bibfnamefont {G.}~\bibnamefont {Xie}}, \bibinfo {author}
  {\bibfnamefont {T.}~\bibnamefont {Wang}},\ and\ \bibinfo {author}
  {\bibfnamefont {G.}~\bibnamefont {Zhang}},\ }\bibfield  {title} {\bibinfo
  {title} {{Thermal Conductivity of Amorphous Materials}},\ }\href@noop {}
  {\bibfield  {journal} {\bibinfo  {journal} {Adv. Funct. Mater.}\ }\textbf
  {\bibinfo {volume} {30}},\ \bibinfo {pages} {1903829} (\bibinfo {year}
  {2019})}\BibitemShut {NoStop}%
\bibitem [{\citenamefont {Jund}\ and\ \citenamefont
  {Jullien}(1999)}]{Jund1999}%
  \BibitemOpen
  \bibfield  {author} {\bibinfo {author} {\bibfnamefont {P.}~\bibnamefont
  {Jund}}\ and\ \bibinfo {author} {\bibfnamefont {R.}~\bibnamefont {Jullien}},\
  }\bibfield  {title} {\bibinfo {title} {Molecular-dynamics calculation of the
  thermal conductivity of vitreous silica},\ }\href
  {https://doi.org/10.1103/PhysRevB.59.13707} {\bibfield  {journal} {\bibinfo
  {journal} {Phys. Rev. B}\ }\textbf {\bibinfo {volume} {59}},\ \bibinfo
  {pages} {13707} (\bibinfo {year} {1999})}\BibitemShut {NoStop}%
\bibitem [{\citenamefont {Shenogin}\ \emph {et~al.}(2009)\citenamefont
  {Shenogin}, \citenamefont {Bodapati}, \citenamefont {Keblinski},\ and\
  \citenamefont {McGaughey}}]{Shenogin2009}%
  \BibitemOpen
  \bibfield  {author} {\bibinfo {author} {\bibfnamefont {S.}~\bibnamefont
  {Shenogin}}, \bibinfo {author} {\bibfnamefont {A.}~\bibnamefont {Bodapati}},
  \bibinfo {author} {\bibfnamefont {P.}~\bibnamefont {Keblinski}},\ and\
  \bibinfo {author} {\bibfnamefont {A.~J.~H.}\ \bibnamefont {McGaughey}},\
  }\bibfield  {title} {\bibinfo {title} {Predicting the thermal conductivity of
  inorganic and polymeric glasses: The role of anharmonicity},\ }\href
  {https://doi.org/10.1063/1.3073954} {\bibfield  {journal} {\bibinfo
  {journal} {J. Appl. Phys.}\ }\textbf {\bibinfo {volume} {105}},\ \bibinfo
  {pages} {034906} (\bibinfo {year} {2009})}\BibitemShut {NoStop}%
\bibitem [{\citenamefont {Sosso}\ \emph {et~al.}(2012)\citenamefont {Sosso},
  \citenamefont {Donadio}, \citenamefont {Caravati}, \citenamefont {Behler},\
  and\ \citenamefont {Bernasconi}}]{sosso2012}%
  \BibitemOpen
  \bibfield  {author} {\bibinfo {author} {\bibfnamefont {G.~C.}\ \bibnamefont
  {Sosso}}, \bibinfo {author} {\bibfnamefont {D.}~\bibnamefont {Donadio}},
  \bibinfo {author} {\bibfnamefont {S.}~\bibnamefont {Caravati}}, \bibinfo
  {author} {\bibfnamefont {J.}~\bibnamefont {Behler}},\ and\ \bibinfo {author}
  {\bibfnamefont {M.}~\bibnamefont {Bernasconi}},\ }\bibfield  {title}
  {\bibinfo {title} {Thermal transport in phase-change materials from atomistic
  simulations},\ }\href@noop {} {\bibfield  {journal} {\bibinfo  {journal}
  {Phys. Rev. B}\ }\textbf {\bibinfo {volume} {86}},\ \bibinfo {pages} {104301}
  (\bibinfo {year} {2012})}\BibitemShut {NoStop}%
\bibitem [{\citenamefont {Allen}\ and\ \citenamefont
  {Feldman}(1989)}]{Allen:1989ua}%
  \BibitemOpen
  \bibfield  {author} {\bibinfo {author} {\bibfnamefont {P.~B.}\ \bibnamefont
  {Allen}}\ and\ \bibinfo {author} {\bibfnamefont {J.~L.}\ \bibnamefont
  {Feldman}},\ }\bibfield  {title} {\bibinfo {title} {{Thermal Conductivity of
  Glasses: Theory and Application to Amorphous Si}},\ }\href@noop {} {\bibfield
   {journal} {\bibinfo  {journal} {Phys. Rev. Lett.}\ }\textbf {\bibinfo
  {volume} {62}},\ \bibinfo {pages} {645} (\bibinfo {year} {1989})}\BibitemShut
  {NoStop}%
\bibitem [{\citenamefont {Allen}\ and\ \citenamefont
  {Feldman}(1993)}]{Allen:1993uc}%
  \BibitemOpen
  \bibfield  {author} {\bibinfo {author} {\bibfnamefont {P.~B.}\ \bibnamefont
  {Allen}}\ and\ \bibinfo {author} {\bibfnamefont {J.~L.}\ \bibnamefont
  {Feldman}},\ }\bibfield  {title} {\bibinfo {title} {{Thermal conductivity of
  disordered harmonic solids}},\ }\href@noop {} {\bibfield  {journal} {\bibinfo
   {journal} {Phys. Rev. B}\ }\textbf {\bibinfo {volume} {48}},\ \bibinfo
  {pages} {12581} (\bibinfo {year} {1993})}\BibitemShut {NoStop}%
\bibitem [{\citenamefont {Feldman}\ \emph {et~al.}(1993)\citenamefont
  {Feldman}, \citenamefont {Kluge}, \citenamefont {Allen},\ and\ \citenamefont
  {Wooten}}]{Feldman:1993tn}%
  \BibitemOpen
  \bibfield  {author} {\bibinfo {author} {\bibfnamefont {J.~L.}\ \bibnamefont
  {Feldman}}, \bibinfo {author} {\bibfnamefont {M.~D.}\ \bibnamefont {Kluge}},
  \bibinfo {author} {\bibfnamefont {P.~B.}\ \bibnamefont {Allen}},\ and\
  \bibinfo {author} {\bibfnamefont {F.}~\bibnamefont {Wooten}},\ }\bibfield
  {title} {\bibinfo {title} {{Thermal conductivity and localization in glasses:
  Numerical study of a model of amorphous silicon}},\ }\href@noop {} {\bibfield
   {journal} {\bibinfo  {journal} {Phys. Rev. B}\ }\textbf {\bibinfo {volume}
  {48}},\ \bibinfo {pages} {12589} (\bibinfo {year} {1993})}\BibitemShut
  {NoStop}%
\bibitem [{\citenamefont {Feldman}\ \emph {et~al.}(1999)\citenamefont
  {Feldman}, \citenamefont {Allen},\ and\ \citenamefont
  {Bickham}}]{Feldman:1999uw}%
  \BibitemOpen
  \bibfield  {author} {\bibinfo {author} {\bibfnamefont {J.~L.}\ \bibnamefont
  {Feldman}}, \bibinfo {author} {\bibfnamefont {P.~B.}\ \bibnamefont {Allen}},\
  and\ \bibinfo {author} {\bibfnamefont {S.~R.}\ \bibnamefont {Bickham}},\
  }\bibfield  {title} {\bibinfo {title} {{Numerical study of low-frequency
  vibrations in amorphous silicon}},\ }\href@noop {} {\bibfield  {journal}
  {\bibinfo  {journal} {Phys. Rev. B}\ }\textbf {\bibinfo {volume} {59}},\
  \bibinfo {pages} {3551} (\bibinfo {year} {1999})}\BibitemShut {NoStop}%
\bibitem [{\citenamefont {Allen}\ \emph {et~al.}(1999)\citenamefont {Allen},
  \citenamefont {Feldman}, \citenamefont {Fabian},\ and\ \citenamefont
  {Wooten}}]{allen1999}%
  \BibitemOpen
  \bibfield  {author} {\bibinfo {author} {\bibfnamefont {P.~B.}\ \bibnamefont
  {Allen}}, \bibinfo {author} {\bibfnamefont {J.~L.}\ \bibnamefont {Feldman}},
  \bibinfo {author} {\bibfnamefont {J.}~\bibnamefont {Fabian}},\ and\ \bibinfo
  {author} {\bibfnamefont {F.}~\bibnamefont {Wooten}},\ }\bibfield  {title}
  {\bibinfo {title} {Diffusons, locons and propagons: Character of atomic
  vibrations in amorphous si},\ }\href@noop {} {\bibfield  {journal} {\bibinfo
  {journal} {Philos. Mag. B}\ }\textbf {\bibinfo {volume} {79}},\ \bibinfo
  {pages} {1715} (\bibinfo {year} {1999})}\BibitemShut {NoStop}%
\bibitem [{\citenamefont {Donadio}\ and\ \citenamefont
  {Galli}(2010)}]{Donadio:2010kp}%
  \BibitemOpen
  \bibfield  {author} {\bibinfo {author} {\bibfnamefont {D.}~\bibnamefont
  {Donadio}}\ and\ \bibinfo {author} {\bibfnamefont {G.}~\bibnamefont
  {Galli}},\ }\bibfield  {title} {\bibinfo {title} {{Temperature Dependence of
  the Thermal Conductivity of Thin Silicon Nanowires}},\ }\href@noop {}
  {\bibfield  {journal} {\bibinfo  {journal} {Nano Lett.}\ }\textbf {\bibinfo
  {volume} {10}},\ \bibinfo {pages} {847} (\bibinfo {year} {2010})}\BibitemShut
  {NoStop}%
\bibitem [{\citenamefont {Zhu}\ and\ \citenamefont
  {Ertekin}(2016{\natexlab{a}})}]{Zhu:2016bi}%
  \BibitemOpen
  \bibfield  {author} {\bibinfo {author} {\bibfnamefont {T.}~\bibnamefont
  {Zhu}}\ and\ \bibinfo {author} {\bibfnamefont {E.}~\bibnamefont {Ertekin}},\
  }\bibfield  {title} {\bibinfo {title} {{Phonons, Localization, and Thermal
  Conductivity of Diamond Nanothreads and Amorphous Graphene}},\ }\href@noop {}
  {\bibfield  {journal} {\bibinfo  {journal} {Nano Lett.}\ }\textbf {\bibinfo
  {volume} {16}},\ \bibinfo {pages} {4763} (\bibinfo {year}
  {2016}{\natexlab{a}})}\BibitemShut {NoStop}%
\bibitem [{\citenamefont {Zhu}\ and\ \citenamefont
  {Ertekin}(2016{\natexlab{b}})}]{Zhu:2016ks}%
  \BibitemOpen
  \bibfield  {author} {\bibinfo {author} {\bibfnamefont {T.}~\bibnamefont
  {Zhu}}\ and\ \bibinfo {author} {\bibfnamefont {E.}~\bibnamefont {Ertekin}},\
  }\bibfield  {title} {\bibinfo {title} {{Generalized
  Debye-Peierls/Allen-Feldman model for the lattice thermal conductivity of
  low-dimensional and disordered materials}},\ }\href@noop {} {\bibfield
  {journal} {\bibinfo  {journal} {Phys. Rev. B}\ }\textbf {\bibinfo {volume}
  {93}},\ \bibinfo {pages} {155414} (\bibinfo {year}
  {2016}{\natexlab{b}})}\BibitemShut {NoStop}%
\bibitem [{\citenamefont {Zhu}\ \emph {et~al.}(2018)\citenamefont {Zhu},
  \citenamefont {Swaminathan-Gopalan}, \citenamefont {Cruse}, \citenamefont
  {Stephani},\ and\ \citenamefont {Ertekin}}]{Zhu:2018kf}%
  \BibitemOpen
  \bibfield  {author} {\bibinfo {author} {\bibfnamefont {T.}~\bibnamefont
  {Zhu}}, \bibinfo {author} {\bibfnamefont {K.}~\bibnamefont
  {Swaminathan-Gopalan}}, \bibinfo {author} {\bibfnamefont {K.~J.}\
  \bibnamefont {Cruse}}, \bibinfo {author} {\bibfnamefont {K.}~\bibnamefont
  {Stephani}},\ and\ \bibinfo {author} {\bibfnamefont {E.}~\bibnamefont
  {Ertekin}},\ }\bibfield  {title} {\bibinfo {title} {{Vibrational Energy
  Transport in Hybrid Ordered/Disordered Nanocomposites: Hybridization and
  Avoided Crossings of Localized and Delocalized Modes}},\ }\href@noop {}
  {\bibfield  {journal} {\bibinfo  {journal} {Adv. Funct. Mater.}\ }\textbf
  {\bibinfo {volume} {28}},\ \bibinfo {pages} {1706268} (\bibinfo {year}
  {2018})}\BibitemShut {NoStop}%
\bibitem [{\citenamefont {Zhou}\ \emph {et~al.}(2020)\citenamefont {Zhou},
  \citenamefont {Li}, \citenamefont {Cheng}, \citenamefont {Ni}, \citenamefont
  {Volz}, \citenamefont {Donadio}, \citenamefont {Xiong}, \citenamefont
  {Zhang},\ and\ \citenamefont {Zhang}}]{Zhou:2020jp}%
  \BibitemOpen
  \bibfield  {author} {\bibinfo {author} {\bibfnamefont {T.}~\bibnamefont
  {Zhou}}, \bibinfo {author} {\bibfnamefont {Z.}~\bibnamefont {Li}}, \bibinfo
  {author} {\bibfnamefont {Y.}~\bibnamefont {Cheng}}, \bibinfo {author}
  {\bibfnamefont {Y.}~\bibnamefont {Ni}}, \bibinfo {author} {\bibfnamefont
  {S.}~\bibnamefont {Volz}}, \bibinfo {author} {\bibfnamefont {D.}~\bibnamefont
  {Donadio}}, \bibinfo {author} {\bibfnamefont {S.}~\bibnamefont {Xiong}},
  \bibinfo {author} {\bibfnamefont {W.}~\bibnamefont {Zhang}},\ and\ \bibinfo
  {author} {\bibfnamefont {X.}~\bibnamefont {Zhang}},\ }\bibfield  {title}
  {\bibinfo {title} {{Thermal transport in amorphous small organic materials: a
  mechanistic study}},\ }\href@noop {} {\bibfield  {journal} {\bibinfo
  {journal} {Phys. Chem. Chem. Phys.}\ }\textbf {\bibinfo {volume} {22}},\
  \bibinfo {pages} {3058} (\bibinfo {year} {2020})}\BibitemShut {NoStop}%
\bibitem [{\citenamefont {He}\ \emph {et~al.}(2011{\natexlab{a}})\citenamefont
  {He}, \citenamefont {Donadio},\ and\ \citenamefont {Galli}}]{he2011}%
  \BibitemOpen
  \bibfield  {author} {\bibinfo {author} {\bibfnamefont {Y.}~\bibnamefont
  {He}}, \bibinfo {author} {\bibfnamefont {D.}~\bibnamefont {Donadio}},\ and\
  \bibinfo {author} {\bibfnamefont {G.}~\bibnamefont {Galli}},\ }\bibfield
  {title} {\bibinfo {title} {Heat transport in amorphous silicon: Interplay
  between morphology and disorder},\ }\href@noop {} {\bibfield  {journal}
  {\bibinfo  {journal} {Appl. Phys. Lett.}\ }\textbf {\bibinfo {volume} {98}},\
  \bibinfo {pages} {144101} (\bibinfo {year} {2011}{\natexlab{a}})}\BibitemShut
  {NoStop}%
\bibitem [{\citenamefont {Larkin}\ and\ \citenamefont
  {McGaughey}(2014)}]{Larkin:2014dc}%
  \BibitemOpen
  \bibfield  {author} {\bibinfo {author} {\bibfnamefont {J.~M.}\ \bibnamefont
  {Larkin}}\ and\ \bibinfo {author} {\bibfnamefont {A.~J.~H.}\ \bibnamefont
  {McGaughey}},\ }\bibfield  {title} {\bibinfo {title} {{Thermal conductivity
  accumulation in amorphous silica and amorphous silicon}},\ }\href@noop {}
  {\bibfield  {journal} {\bibinfo  {journal} {Phys. Rev. B}\ }\textbf {\bibinfo
  {volume} {89}},\ \bibinfo {pages} {144303} (\bibinfo {year}
  {2014})}\BibitemShut {NoStop}%
\bibitem [{\citenamefont {Lv}\ and\ \citenamefont
  {Henry}(2016)}]{lv2016direct}%
  \BibitemOpen
  \bibfield  {author} {\bibinfo {author} {\bibfnamefont {W.}~\bibnamefont
  {Lv}}\ and\ \bibinfo {author} {\bibfnamefont {A.}~\bibnamefont {Henry}},\
  }\bibfield  {title} {\bibinfo {title} {Direct calculation of modal
  contributions to thermal conductivity via green--kubo modal analysis},\
  }\href@noop {} {\bibfield  {journal} {\bibinfo  {journal} {New J. Phys.}\
  }\textbf {\bibinfo {volume} {18}},\ \bibinfo {pages} {013028} (\bibinfo
  {year} {2016})}\BibitemShut {NoStop}%
\bibitem [{\citenamefont {Braun}\ \emph {et~al.}(2016)\citenamefont {Braun},
  \citenamefont {Baker}, \citenamefont {Giri}, \citenamefont {Elahi},
  \citenamefont {Artyushkova}, \citenamefont {Beechem}, \citenamefont {Norris},
  \citenamefont {Leseman}, \citenamefont {Gaskins},\ and\ \citenamefont
  {Hopkins}}]{braun2016size}%
  \BibitemOpen
  \bibfield  {author} {\bibinfo {author} {\bibfnamefont {J.~L.}\ \bibnamefont
  {Braun}}, \bibinfo {author} {\bibfnamefont {C.~H.}\ \bibnamefont {Baker}},
  \bibinfo {author} {\bibfnamefont {A.}~\bibnamefont {Giri}}, \bibinfo {author}
  {\bibfnamefont {M.}~\bibnamefont {Elahi}}, \bibinfo {author} {\bibfnamefont
  {K.}~\bibnamefont {Artyushkova}}, \bibinfo {author} {\bibfnamefont {T.~E.}\
  \bibnamefont {Beechem}}, \bibinfo {author} {\bibfnamefont {P.~M.}\
  \bibnamefont {Norris}}, \bibinfo {author} {\bibfnamefont {Z.~C.}\
  \bibnamefont {Leseman}}, \bibinfo {author} {\bibfnamefont {J.~T.}\
  \bibnamefont {Gaskins}},\ and\ \bibinfo {author} {\bibfnamefont {P.~E.}\
  \bibnamefont {Hopkins}},\ }\bibfield  {title} {\bibinfo {title} {Size effects
  on the thermal conductivity of amorphous silicon thin films},\ }\href@noop {}
  {\bibfield  {journal} {\bibinfo  {journal} {Phys. Rev. B}\ }\textbf {\bibinfo
  {volume} {93}},\ \bibinfo {pages} {140201(R)} (\bibinfo {year}
  {2016})}\BibitemShut {NoStop}%
\bibitem [{\citenamefont {Moon}\ \emph {et~al.}(2018)\citenamefont {Moon},
  \citenamefont {Latour},\ and\ \citenamefont {Minnich}}]{moon2018}%
  \BibitemOpen
  \bibfield  {author} {\bibinfo {author} {\bibfnamefont {J.}~\bibnamefont
  {Moon}}, \bibinfo {author} {\bibfnamefont {B.}~\bibnamefont {Latour}},\ and\
  \bibinfo {author} {\bibfnamefont {A.~J.}\ \bibnamefont {Minnich}},\
  }\bibfield  {title} {\bibinfo {title} {Propagating elastic vibrations
  dominate thermal conduction in amorphous silicon},\ }\href@noop {} {\bibfield
   {journal} {\bibinfo  {journal} {Phys. Rev. B}\ }\textbf {\bibinfo {volume}
  {97}},\ \bibinfo {pages} {024201} (\bibinfo {year} {2018})}\BibitemShut
  {NoStop}%
\bibitem [{\citenamefont {Simoncelli}\ \emph {et~al.}(2019)\citenamefont
  {Simoncelli}, \citenamefont {Marzari},\ and\ \citenamefont
  {Mauri}}]{simoncelli2019}%
  \BibitemOpen
  \bibfield  {author} {\bibinfo {author} {\bibfnamefont {M.}~\bibnamefont
  {Simoncelli}}, \bibinfo {author} {\bibfnamefont {N.}~\bibnamefont
  {Marzari}},\ and\ \bibinfo {author} {\bibfnamefont {F.}~\bibnamefont
  {Mauri}},\ }\bibfield  {title} {\bibinfo {title} {Unified theory of thermal
  transport in crystals and glasses},\ }\href@noop {} {\bibfield  {journal}
  {\bibinfo  {journal} {Nat. Phys.}\ }\textbf {\bibinfo {volume} {15}},\
  \bibinfo {pages} {809} (\bibinfo {year} {2019})}\BibitemShut {NoStop}%
\bibitem [{\citenamefont {Isaeva}\ \emph {et~al.}(2019)\citenamefont {Isaeva},
  \citenamefont {Barbalinardo}, \citenamefont {Donadio},\ and\ \citenamefont
  {Baroni}}]{isaeva2019}%
  \BibitemOpen
  \bibfield  {author} {\bibinfo {author} {\bibfnamefont {L.}~\bibnamefont
  {Isaeva}}, \bibinfo {author} {\bibfnamefont {G.}~\bibnamefont
  {Barbalinardo}}, \bibinfo {author} {\bibfnamefont {D.}~\bibnamefont
  {Donadio}},\ and\ \bibinfo {author} {\bibfnamefont {S.}~\bibnamefont
  {Baroni}},\ }\bibfield  {title} {\bibinfo {title} {Modeling heat transport in
  crystals and glasses from a unified lattice-dynamical approach},\ }\href@noop
  {} {\bibfield  {journal} {\bibinfo  {journal} {Nat. Comm.}\ }\textbf
  {\bibinfo {volume} {10}},\ \bibinfo {pages} {1} (\bibinfo {year}
  {2019})}\BibitemShut {NoStop}%
\bibitem [{\citenamefont {Neogi}\ and\ \citenamefont
  {Donadio}(2020)}]{Neogi:2020vz}%
  \BibitemOpen
  \bibfield  {author} {\bibinfo {author} {\bibfnamefont {S.}~\bibnamefont
  {Neogi}}\ and\ \bibinfo {author} {\bibfnamefont {D.}~\bibnamefont
  {Donadio}},\ }\bibfield  {title} {\bibinfo {title} {{Anisotropic In-Plane
  Phonon Transport in Silicon Membranes Guided by Nanoscale Surface
  Resonators}},\ }\href@noop {} {\bibfield  {journal} {\bibinfo  {journal}
  {Phys. Rev. Appl.}\ }\textbf {\bibinfo {volume} {14}},\ \bibinfo {pages}
  {024004} (\bibinfo {year} {2020})}\BibitemShut {NoStop}%
\bibitem [{\citenamefont {Zink}\ \emph {et~al.}(2006)\citenamefont {Zink},
  \citenamefont {Pietri},\ and\ \citenamefont {Hellman}}]{Zink:2006wt}%
  \BibitemOpen
  \bibfield  {author} {\bibinfo {author} {\bibfnamefont {B.~L.}\ \bibnamefont
  {Zink}}, \bibinfo {author} {\bibfnamefont {R.}~\bibnamefont {Pietri}},\ and\
  \bibinfo {author} {\bibfnamefont {F.}~\bibnamefont {Hellman}},\ }\bibfield
  {title} {\bibinfo {title} {{Thermal conductivity and specific heat of
  thin-film amorphous silicon}},\ }\href@noop {} {\bibfield  {journal}
  {\bibinfo  {journal} {Phys. Rev. Lett.}\ }\textbf {\bibinfo {volume} {96}},\
  \bibinfo {pages} {55902} (\bibinfo {year} {2006})}\BibitemShut {NoStop}%
\bibitem [{\citenamefont {Abeles}(1963)}]{Abeles:1963}%
  \BibitemOpen
  \bibfield  {author} {\bibinfo {author} {\bibfnamefont {B.}~\bibnamefont
  {Abeles}},\ }\bibfield  {title} {\bibinfo {title} {Lattice thermal
  conductivity of disordered semiconductor alloys at high temperatures},\
  }\href {https://doi.org/10.1103/PhysRev.131.1906} {\bibfield  {journal}
  {\bibinfo  {journal} {Phys. Rev.}\ }\textbf {\bibinfo {volume} {131}},\
  \bibinfo {pages} {1906} (\bibinfo {year} {1963})}\BibitemShut {NoStop}%
\bibitem [{\citenamefont {Garg}\ \emph {et~al.}(2011)\citenamefont {Garg},
  \citenamefont {Bonini}, \citenamefont {Kozinsky},\ and\ \citenamefont
  {Marzari}}]{Garg:2011hi}%
  \BibitemOpen
  \bibfield  {author} {\bibinfo {author} {\bibfnamefont {J.}~\bibnamefont
  {Garg}}, \bibinfo {author} {\bibfnamefont {N.}~\bibnamefont {Bonini}},
  \bibinfo {author} {\bibfnamefont {B.}~\bibnamefont {Kozinsky}},\ and\
  \bibinfo {author} {\bibfnamefont {N.}~\bibnamefont {Marzari}},\ }\bibfield
  {title} {\bibinfo {title} {{Role of Disorder and Anharmonicity in the Thermal
  Conductivity of Silicon-Germanium Alloys: A First-Principles Study}},\
  }\href@noop {} {\bibfield  {journal} {\bibinfo  {journal} {Phys. Rev. Lett.}\
  }\textbf {\bibinfo {volume} {106}},\ \bibinfo {pages} {045901} (\bibinfo
  {year} {2011})}\BibitemShut {NoStop}%
\bibitem [{\citenamefont {Wang}\ and\ \citenamefont
  {Mingo}(2010)}]{Wang:2010hw}%
  \BibitemOpen
  \bibfield  {author} {\bibinfo {author} {\bibfnamefont {Z.}~\bibnamefont
  {Wang}}\ and\ \bibinfo {author} {\bibfnamefont {N.}~\bibnamefont {Mingo}},\
  }\bibfield  {title} {\bibinfo {title} {{Diameter dependence of SiGe nanowire
  thermal conductivity}},\ }\href@noop {} {\bibfield  {journal} {\bibinfo
  {journal} {Appl. Phys. Lett.}\ }\textbf {\bibinfo {volume} {97}},\ \bibinfo
  {pages} {101903} (\bibinfo {year} {2010})}\BibitemShut {NoStop}%
\bibitem [{\citenamefont {Larkin}\ and\ \citenamefont
  {McGaughey}(2013)}]{larkin2013predicting}%
  \BibitemOpen
  \bibfield  {author} {\bibinfo {author} {\bibfnamefont {J.~M.}\ \bibnamefont
  {Larkin}}\ and\ \bibinfo {author} {\bibfnamefont {A.~J.~H.}\ \bibnamefont
  {McGaughey}},\ }\bibfield  {title} {\bibinfo {title} {Predicting alloy
  vibrational mode properties using lattice dynamics calculations, molecular
  dynamics simulations, and the virtual crystal approximation},\ }\href@noop {}
  {\bibfield  {journal} {\bibinfo  {journal} {J. Appl. Phys.}\ }\textbf
  {\bibinfo {volume} {114}},\ \bibinfo {pages} {023507} (\bibinfo {year}
  {2013})}\BibitemShut {NoStop}%
\bibitem [{\citenamefont {Aksamija}\ and\ \citenamefont
  {Knezevic}(2013)}]{Aksamija:2013hd}%
  \BibitemOpen
  \bibfield  {author} {\bibinfo {author} {\bibfnamefont {Z.}~\bibnamefont
  {Aksamija}}\ and\ \bibinfo {author} {\bibfnamefont {I.}~\bibnamefont
  {Knezevic}},\ }\bibfield  {title} {\bibinfo {title} {{Thermal conductivity of
  Si 1{\textminus}xGe x/Si 1{\textminus}yGe ysuperlattices: Competition between
  interfacial and internal scattering}},\ }\href@noop {} {\bibfield  {journal}
  {\bibinfo  {journal} {Phys. Rev. B}\ }\textbf {\bibinfo {volume} {88}},\
  \bibinfo {pages} {746} (\bibinfo {year} {2013})}\BibitemShut {NoStop}%
\bibitem [{\citenamefont {Upadhyaya}\ \emph {et~al.}(2015)\citenamefont
  {Upadhyaya}, \citenamefont {Khatami},\ and\ \citenamefont
  {Aksamija}}]{Upadhyaya:2015ir}%
  \BibitemOpen
  \bibfield  {author} {\bibinfo {author} {\bibfnamefont {M.}~\bibnamefont
  {Upadhyaya}}, \bibinfo {author} {\bibfnamefont {S.~N.}\ \bibnamefont
  {Khatami}},\ and\ \bibinfo {author} {\bibfnamefont {Z.}~\bibnamefont
  {Aksamija}},\ }\bibfield  {title} {\bibinfo {title} {{Engineering thermal
  transport in SiGe-based nanostructures for thermoelectric applications}},\
  }\href@noop {} {\bibfield  {journal} {\bibinfo  {journal} {J. Mater. Res.}\
  }\textbf {\bibinfo {volume} {30}},\ \bibinfo {pages} {2649} (\bibinfo {year}
  {2015})}\BibitemShut {NoStop}%
\bibitem [{\citenamefont {Khatami}\ and\ \citenamefont
  {Aksamija}(2016)}]{PhysRevApplied.6.014015}%
  \BibitemOpen
  \bibfield  {author} {\bibinfo {author} {\bibfnamefont {S.~N.}\ \bibnamefont
  {Khatami}}\ and\ \bibinfo {author} {\bibfnamefont {Z.}~\bibnamefont
  {Aksamija}},\ }\bibfield  {title} {\bibinfo {title} {{Lattice Thermal
  Conductivity of the Binary and Ternary Group-IV Alloys Si-Sn, Ge-Sn, and
  Si-Ge-Sn}},\ }\href@noop {} {\bibfield  {journal} {\bibinfo  {journal} {Phys.
  Rev. Applied}\ }\textbf {\bibinfo {volume} {6}},\ \bibinfo {pages} {014015}
  (\bibinfo {year} {2016})}\BibitemShut {NoStop}%
\bibitem [{\citenamefont {Xiong}\ \emph {et~al.}(2017)\citenamefont {Xiong},
  \citenamefont {Selli}, \citenamefont {Neogi},\ and\ \citenamefont
  {Donadio}}]{Xiong:2017if}%
  \BibitemOpen
  \bibfield  {author} {\bibinfo {author} {\bibfnamefont {S.}~\bibnamefont
  {Xiong}}, \bibinfo {author} {\bibfnamefont {D.}~\bibnamefont {Selli}},
  \bibinfo {author} {\bibfnamefont {S.}~\bibnamefont {Neogi}},\ and\ \bibinfo
  {author} {\bibfnamefont {D.}~\bibnamefont {Donadio}},\ }\bibfield  {title}
  {\bibinfo {title} {{Native surface oxide turns alloyed silicon membranes into
  nanophononic metamaterials with ultralow thermal conductivity}},\ }\href@noop
  {} {\bibfield  {journal} {\bibinfo  {journal} {Phys. Rev. B}\ }\textbf
  {\bibinfo {volume} {95}},\ \bibinfo {pages} {180301(R)} (\bibinfo {year}
  {2017})}\BibitemShut {NoStop}%
\bibitem [{\citenamefont {Ferrando-Villalba}\ \emph {et~al.}(2020)\citenamefont
  {Ferrando-Villalba}, \citenamefont {Chen}, \citenamefont {Lopeand{\'\i}a},
  \citenamefont {Alvarez}, \citenamefont {Alonso}, \citenamefont {Garriga},
  \citenamefont {Santiso}, \citenamefont {Garcia}, \citenamefont {Go{\~n}i},
  \citenamefont {Donadio},\ and\ \citenamefont
  {Rodriguez-Viejo}}]{FerrandoVillalba:2020hc}%
  \BibitemOpen
  \bibfield  {author} {\bibinfo {author} {\bibfnamefont {P.}~\bibnamefont
  {Ferrando-Villalba}}, \bibinfo {author} {\bibfnamefont {S.}~\bibnamefont
  {Chen}}, \bibinfo {author} {\bibfnamefont {A.~F.}\ \bibnamefont
  {Lopeand{\'\i}a}}, \bibinfo {author} {\bibfnamefont {F.~X.}\ \bibnamefont
  {Alvarez}}, \bibinfo {author} {\bibfnamefont {M.~I.}\ \bibnamefont {Alonso}},
  \bibinfo {author} {\bibfnamefont {M.}~\bibnamefont {Garriga}}, \bibinfo
  {author} {\bibfnamefont {J.}~\bibnamefont {Santiso}}, \bibinfo {author}
  {\bibfnamefont {G.}~\bibnamefont {Garcia}}, \bibinfo {author} {\bibfnamefont
  {A.~R.}\ \bibnamefont {Go{\~n}i}}, \bibinfo {author} {\bibfnamefont
  {D.}~\bibnamefont {Donadio}},\ and\ \bibinfo {author} {\bibfnamefont
  {J.}~\bibnamefont {Rodriguez-Viejo}},\ }\bibfield  {title} {\bibinfo {title}
  {{Beating the thermal conductivity alloy limit using long-period
  compositionally graded Si1-xGex superlattices}},\ }\href@noop {} {\bibfield
  {journal} {\bibinfo  {journal} {J. Phys. Chem. C}\textbf {\bibinfo {volume} {124}},\ \bibinfo {pages}
  {19864-19872}} (\bibinfo {year} {2020})}\BibitemShut {NoStop}%
\bibitem [{\citenamefont {Mondal}\ \emph {et~al.}(2017)\citenamefont {Mondal},
  \citenamefont {Vidhyadhiraja}, \citenamefont {Berlijn}, \citenamefont
  {Moreno},\ and\ \citenamefont {Jarrell}}]{Mondal:2017du}%
  \BibitemOpen
  \bibfield  {author} {\bibinfo {author} {\bibfnamefont {W.~R.}\ \bibnamefont
  {Mondal}}, \bibinfo {author} {\bibfnamefont {N.~S.}\ \bibnamefont
  {Vidhyadhiraja}}, \bibinfo {author} {\bibfnamefont {T.}~\bibnamefont
  {Berlijn}}, \bibinfo {author} {\bibfnamefont {J.}~\bibnamefont {Moreno}},\
  and\ \bibinfo {author} {\bibfnamefont {M.}~\bibnamefont {Jarrell}},\
  }\bibfield  {title} {\bibinfo {title} {Localization of phonons in
  mass-disordered alloys: A typical medium dynamical cluster approach},\ }\href
  {https://doi.org/10.1103/PhysRevB.96.014203} {\bibfield  {journal} {\bibinfo
  {journal} {Phys. Rev. B}\ }\textbf {\bibinfo {volume} {96}},\ \bibinfo
  {pages} {014203} (\bibinfo {year} {2017})}\BibitemShut {NoStop}%
\bibitem [{\citenamefont {Giri}\ \emph {et~al.}(2018)\citenamefont {Giri},
  \citenamefont {Donovan},\ and\ \citenamefont
  {Hopkins}}]{giri2018localization}%
  \BibitemOpen
  \bibfield  {author} {\bibinfo {author} {\bibfnamefont {A.}~\bibnamefont
  {Giri}}, \bibinfo {author} {\bibfnamefont {B.~F.}\ \bibnamefont {Donovan}},\
  and\ \bibinfo {author} {\bibfnamefont {P.~E.}\ \bibnamefont {Hopkins}},\
  }\bibfield  {title} {\bibinfo {title} {Localization of vibrational modes
  leads to reduced thermal conductivity of amorphous heterostructures},\
  }\href@noop {} {\bibfield  {journal} {\bibinfo  {journal} {Phys. Rev.
  Mater.}\ }\textbf {\bibinfo {volume} {2}},\ \bibinfo {pages} {056002}
  (\bibinfo {year} {2018})}\BibitemShut {NoStop}%
\bibitem [{\citenamefont {Tersoff}(1989)}]{tersoff1989modeling}%
  \BibitemOpen
  \bibfield  {author} {\bibinfo {author} {\bibfnamefont {J.}~\bibnamefont
  {Tersoff}},\ }\bibfield  {title} {\bibinfo {title} {Modeling solid-state
  chemistry: Interatomic potentials for multicomponent systems},\ }\href@noop
  {} {\bibfield  {journal} {\bibinfo  {journal} {Phys. Rev. B}\ }\textbf
  {\bibinfo {volume} {39}},\ \bibinfo {pages} {5566} (\bibinfo {year}
  {1989})}\BibitemShut {NoStop}%
\bibitem [{\citenamefont {Plimpton}(1993)}]{plimpton1993fast}%
  \BibitemOpen
  \bibfield  {author} {\bibinfo {author} {\bibfnamefont {S.}~\bibnamefont
  {Plimpton}},\ }\href@noop {} {\emph {\bibinfo {title} {Fast parallel
  algorithms for short-range molecular dynamics}}},\ \bibinfo {type} {Tech.
  Rep.}\ (\bibinfo  {institution} {Sandia National Labs., Albuquerque, NM
  (United States)},\ \bibinfo {year} {1993})\BibitemShut {NoStop}%
\bibitem [{\citenamefont {Bussi}\ \emph {et~al.}(2007)\citenamefont {Bussi},
  \citenamefont {Donadio},\ and\ \citenamefont
  {Parrinello}}]{bussi2007canonical}%
  \BibitemOpen
  \bibfield  {author} {\bibinfo {author} {\bibfnamefont {G.}~\bibnamefont
  {Bussi}}, \bibinfo {author} {\bibfnamefont {D.}~\bibnamefont {Donadio}},\
  and\ \bibinfo {author} {\bibfnamefont {M.}~\bibnamefont {Parrinello}},\
  }\bibfield  {title} {\bibinfo {title} {Canonical sampling through velocity
  rescaling},\ }\href@noop {} {\bibfield  {journal} {\bibinfo  {journal} {J.
  Chem. Phys.}\ }\textbf {\bibinfo {volume} {126}},\ \bibinfo {pages} {014101}
  (\bibinfo {year} {2007})}\BibitemShut {NoStop}%
\bibitem [{\citenamefont {Martyna}\ \emph {et~al.}(1992)\citenamefont
  {Martyna}, \citenamefont {Klein},\ and\ \citenamefont
  {Tuckerman}}]{Martyna:1992gy}%
  \BibitemOpen
  \bibfield  {author} {\bibinfo {author} {\bibfnamefont {G.~J.}\ \bibnamefont
  {Martyna}}, \bibinfo {author} {\bibfnamefont {M.~L.}\ \bibnamefont {Klein}},\
  and\ \bibinfo {author} {\bibfnamefont {M.}~\bibnamefont {Tuckerman}},\
  }\bibfield  {title} {\bibinfo {title} {{Nos{\'e}{\textendash}Hoover chains:
  The canonical ensemble via continuous dynamics}},\ }\href@noop {} {\bibfield
  {journal} {\bibinfo  {journal} {J. Chem. Phys.}\ }\textbf {\bibinfo {volume}
  {97}},\ \bibinfo {pages} {2635} (\bibinfo {year} {1992})}\BibitemShut
  {NoStop}%
\bibitem [{\citenamefont {Ishimaru}\ \emph {et~al.}(1997)\citenamefont
  {Ishimaru}, \citenamefont {Munetoh},\ and\ \citenamefont
  {Motooka}}]{ishimaru1997generation}%
  \BibitemOpen
  \bibfield  {author} {\bibinfo {author} {\bibfnamefont {M.}~\bibnamefont
  {Ishimaru}}, \bibinfo {author} {\bibfnamefont {S.}~\bibnamefont {Munetoh}},\
  and\ \bibinfo {author} {\bibfnamefont {T.}~\bibnamefont {Motooka}},\
  }\bibfield  {title} {\bibinfo {title} {Generation of amorphous silicon
  structures by rapid quenching: A molecular-dynamics study},\ }\href@noop {}
  {\bibfield  {journal} {\bibinfo  {journal} {Phys. Rev. B}\ }\textbf {\bibinfo
  {volume} {56}},\ \bibinfo {pages} {15133} (\bibinfo {year}
  {1997})}\BibitemShut {NoStop}%
\bibitem [{\citenamefont {Sheppard}\ \emph {et~al.}(2008)\citenamefont
  {Sheppard}, \citenamefont {Terrell},\ and\ \citenamefont
  {Henkelman}}]{sheppard2008optimization}%
  \BibitemOpen
  \bibfield  {author} {\bibinfo {author} {\bibfnamefont {D.}~\bibnamefont
  {Sheppard}}, \bibinfo {author} {\bibfnamefont {R.}~\bibnamefont {Terrell}},\
  and\ \bibinfo {author} {\bibfnamefont {G.}~\bibnamefont {Henkelman}},\
  }\bibfield  {title} {\bibinfo {title} {Optimization methods for finding
  minimum energy paths},\ }\href@noop {} {\bibfield  {journal} {\bibinfo
  {journal} {J. Chem. Phys.}\ }\textbf {\bibinfo {volume} {128}},\ \bibinfo
  {pages} {134106} (\bibinfo {year} {2008})}\BibitemShut {NoStop}%
\bibitem [{\citenamefont {Zwanzig}(1965)}]{ZWANZIG:1965tp}%
  \BibitemOpen
  \bibfield  {author} {\bibinfo {author} {\bibfnamefont {R.}~\bibnamefont
  {Zwanzig}},\ }\bibfield  {title} {\bibinfo {title} {{Time-Correlation
  Functions and Transport Coefficients in Statistical Mechanics}},\ }\href@noop
  {} {\bibfield  {journal} {\bibinfo  {journal} {Annu. Rev. Phys. Chem}\
  }\textbf {\bibinfo {volume} {16}},\ \bibinfo {pages} {67} (\bibinfo {year}
  {1965})}\BibitemShut {NoStop}%
\bibitem [{\citenamefont {Ercole}\ \emph {et~al.}(2017)\citenamefont {Ercole},
  \citenamefont {Marcolongo},\ and\ \citenamefont {Baroni}}]{Ercole:2017eea}%
  \BibitemOpen
  \bibfield  {author} {\bibinfo {author} {\bibfnamefont {L.}~\bibnamefont
  {Ercole}}, \bibinfo {author} {\bibfnamefont {A.}~\bibnamefont {Marcolongo}},\
  and\ \bibinfo {author} {\bibfnamefont {S.}~\bibnamefont {Baroni}},\
  }\bibfield  {title} {\bibinfo {title} {{Accurate thermal conductivities from
  optimally short molecular dynamics simulations}},\ }\href@noop {} {\bibfield
  {journal} {\bibinfo  {journal} {Sci. Rep.}\ }\textbf {\bibinfo {volume}
  {7}},\ \bibinfo {pages} {15835} (\bibinfo {year} {2017})}\BibitemShut
  {NoStop}%
\bibitem [{\citenamefont {Fan}\ \emph {et~al.}(2017)\citenamefont {Fan},
  \citenamefont {Chen}, \citenamefont {Vierimaa},\ and\ \citenamefont
  {Harju}}]{fan2017efficient}%
  \BibitemOpen
  \bibfield  {author} {\bibinfo {author} {\bibfnamefont {Z.}~\bibnamefont
  {Fan}}, \bibinfo {author} {\bibfnamefont {W.}~\bibnamefont {Chen}}, \bibinfo
  {author} {\bibfnamefont {V.}~\bibnamefont {Vierimaa}},\ and\ \bibinfo
  {author} {\bibfnamefont {A.}~\bibnamefont {Harju}},\ }\bibfield  {title}
  {\bibinfo {title} {Efficient molecular dynamics simulations with many-body
  potentials on graphics processing units},\ }\href@noop {} {\bibfield
  {journal} {\bibinfo  {journal} {Comput. Phys. Commun.}\ }\textbf {\bibinfo
  {volume} {218}},\ \bibinfo {pages} {10} (\bibinfo {year} {2017})}\BibitemShut
  {NoStop}%
\bibitem [{\citenamefont {Fan}\ \emph {et~al.}(2015)\citenamefont {Fan},
  \citenamefont {Pereira}, \citenamefont {Wang}, \citenamefont {Zheng},
  \citenamefont {Donadio},\ and\ \citenamefont {Harju}}]{Fan:2015ba}%
  \BibitemOpen
  \bibfield  {author} {\bibinfo {author} {\bibfnamefont {Z.}~\bibnamefont
  {Fan}}, \bibinfo {author} {\bibfnamefont {L.~F.~C.}\ \bibnamefont {Pereira}},
  \bibinfo {author} {\bibfnamefont {H.-Q.}\ \bibnamefont {Wang}}, \bibinfo
  {author} {\bibfnamefont {J.-C.}\ \bibnamefont {Zheng}}, \bibinfo {author}
  {\bibfnamefont {D.}~\bibnamefont {Donadio}},\ and\ \bibinfo {author}
  {\bibfnamefont {A.}~\bibnamefont {Harju}},\ }\bibfield  {title} {\bibinfo
  {title} {{Force and heat current formulas for many-body potentials in
  molecular dynamics simulations with applications to thermal conductivity
  calculations}},\ }\href@noop {} {\bibfield  {journal} {\bibinfo  {journal}
  {Phys. Rev. B}\ }\textbf {\bibinfo {volume} {92}},\ \bibinfo {pages} {094301}
  (\bibinfo {year} {2015})}\BibitemShut {NoStop}%
\bibitem [{\citenamefont {Barbalinardo}\ \emph {et~al.}(2020)\citenamefont
  {Barbalinardo}, \citenamefont {Chen}, \citenamefont {Lundgren},\ and\
  \citenamefont {Donadio}}]{barbalinardo2020efficient}%
  \BibitemOpen
  \bibfield  {author} {\bibinfo {author} {\bibfnamefont {G.}~\bibnamefont
  {Barbalinardo}}, \bibinfo {author} {\bibfnamefont {Z.}~\bibnamefont {Chen}},
  \bibinfo {author} {\bibfnamefont {N.~W.}\ \bibnamefont {Lundgren}},\ and\
  \bibinfo {author} {\bibfnamefont {D.}~\bibnamefont {Donadio}},\ }\bibfield
  {title} {\bibinfo {title} {Efficient anharmonic lattice dynamics calculations
  of thermal transport in crystalline and disordered solids},\ }\href@noop {}
  {\bibfield  {journal} {\bibinfo  {journal} {Journal of Applied Physics}\
  }\textbf {\bibinfo {volume} {128}},\ \bibinfo {pages} {135104} (\bibinfo
  {year} {2020})}\BibitemShut {NoStop}%
\bibitem [{\citenamefont {Fabian}\ \emph {et~al.}(2003)\citenamefont {Fabian},
  \citenamefont {Feldman}, \citenamefont {Hellberg},\ and\ \citenamefont
  {Nakhmanson}}]{Fabian:2003fv}%
  \BibitemOpen
  \bibfield  {author} {\bibinfo {author} {\bibfnamefont {J.}~\bibnamefont
  {Fabian}}, \bibinfo {author} {\bibfnamefont {J.~L.}\ \bibnamefont {Feldman}},
  \bibinfo {author} {\bibfnamefont {C.~S.}\ \bibnamefont {Hellberg}},\ and\
  \bibinfo {author} {\bibfnamefont {S.~M.}\ \bibnamefont {Nakhmanson}},\
  }\bibfield  {title} {\bibinfo {title} {Numerical study of anharmonic
  vibrational decay in amorphous and paracrystalline silicon},\ }\href
  {https://doi.org/10.1103/PhysRevB.67.224302} {\bibfield  {journal} {\bibinfo
  {journal} {Phys. Rev. B}\ }\textbf {\bibinfo {volume} {67}},\ \bibinfo
  {pages} {224302} (\bibinfo {year} {2003})}\BibitemShut {NoStop}%
\bibitem [{\citenamefont {Cahill}\ \emph {et~al.}(1994)\citenamefont {Cahill},
  \citenamefont {Katiyar},\ and\ \citenamefont {Abelson}}]{Cahill:1994}%
  \BibitemOpen
  \bibfield  {author} {\bibinfo {author} {\bibfnamefont {D.~G.}\ \bibnamefont
  {Cahill}}, \bibinfo {author} {\bibfnamefont {M.}~\bibnamefont {Katiyar}},\
  and\ \bibinfo {author} {\bibfnamefont {J.~R.}\ \bibnamefont {Abelson}},\
  }\bibfield  {title} {\bibinfo {title} {Thermal conductivity of a-si:h thin
  films},\ }\href {https://doi.org/10.1103/PhysRevB.50.6077} {\bibfield
  {journal} {\bibinfo  {journal} {Phys. Rev. B}\ }\textbf {\bibinfo {volume}
  {50}},\ \bibinfo {pages} {6077} (\bibinfo {year} {1994})}\BibitemShut
  {NoStop}%
\bibitem [{\citenamefont {Schober}\ and\ \citenamefont
  {Oligschleger}(1996)}]{schober1996low}%
  \BibitemOpen
  \bibfield  {author} {\bibinfo {author} {\bibfnamefont {H.~R.}\ \bibnamefont
  {Schober}}\ and\ \bibinfo {author} {\bibfnamefont {C.}~\bibnamefont
  {Oligschleger}},\ }\bibfield  {title} {\bibinfo {title} {Low-frequency
  vibrations in a model glass},\ }\href@noop {} {\bibfield  {journal} {\bibinfo
   {journal} {Phys. Rev. B}\ }\textbf {\bibinfo {volume} {53}},\ \bibinfo
  {pages} {11469} (\bibinfo {year} {1996})}\BibitemShut {NoStop}%
\bibitem [{\citenamefont {Shevchik}\ \emph {et~al.}(1973)\citenamefont
  {Shevchik}, \citenamefont {Lannin},\ and\ \citenamefont
  {Tejeda}}]{shevchik1973}%
  \BibitemOpen
  \bibfield  {author} {\bibinfo {author} {\bibfnamefont {N.~J.}\ \bibnamefont
  {Shevchik}}, \bibinfo {author} {\bibfnamefont {J.~S.}\ \bibnamefont
  {Lannin}},\ and\ \bibinfo {author} {\bibfnamefont {J.}~\bibnamefont
  {Tejeda}},\ }\bibfield  {title} {\bibinfo {title} {Structure and raman
  scattering of amorphous ge 0.5 si 0.5},\ }\href@noop {} {\bibfield  {journal}
  {\bibinfo  {journal} {Phys. Rev. B}\ }\textbf {\bibinfo {volume} {7}},\
  \bibinfo {pages} {3987} (\bibinfo {year} {1973})}\BibitemShut {NoStop}%
\bibitem [{\citenamefont {Kugler}\ \emph {et~al.}(1989)\citenamefont {Kugler},
  \citenamefont {Moln{\'a}r}, \citenamefont {Pet{\"o}}, \citenamefont
  {Zsoldos}, \citenamefont {Rosta}, \citenamefont {Menelle},\ and\
  \citenamefont {Bellissent}}]{kugler1989}%
  \BibitemOpen
  \bibfield  {author} {\bibinfo {author} {\bibfnamefont {S.}~\bibnamefont
  {Kugler}}, \bibinfo {author} {\bibfnamefont {G.}~\bibnamefont {Moln{\'a}r}},
  \bibinfo {author} {\bibfnamefont {G.}~\bibnamefont {Pet{\"o}}}, \bibinfo
  {author} {\bibfnamefont {E.}~\bibnamefont {Zsoldos}}, \bibinfo {author}
  {\bibfnamefont {L.}~\bibnamefont {Rosta}}, \bibinfo {author} {\bibfnamefont
  {A.}~\bibnamefont {Menelle}},\ and\ \bibinfo {author} {\bibfnamefont
  {R.}~\bibnamefont {Bellissent}},\ }\bibfield  {title} {\bibinfo {title}
  {Neutron-diffraction study of the structure of evaporated pure amorphous
  silicon},\ }\href@noop {} {\bibfield  {journal} {\bibinfo  {journal} {Phys.
  Rev. B}\ }\textbf {\bibinfo {volume} {40}},\ \bibinfo {pages} {8030}
  (\bibinfo {year} {1989})}\BibitemShut {NoStop}%
\bibitem [{\citenamefont {Graczyk}\ and\ \citenamefont
  {Chaudhari}(1973)}]{graczyk1973scanning}%
  \BibitemOpen
  \bibfield  {author} {\bibinfo {author} {\bibfnamefont {J.}~\bibnamefont
  {Graczyk}}\ and\ \bibinfo {author} {\bibfnamefont {P.}~\bibnamefont
  {Chaudhari}},\ }\bibfield  {title} {\bibinfo {title} {A scanning electron
  diffraction study of vapor-deposited and ion implanted thin films of ge
  (i)},\ }\href@noop {} {\bibfield  {journal} {\bibinfo  {journal} {Phys.
  Status Solidi B}\ }\textbf {\bibinfo {volume} {58}},\ \bibinfo {pages} {163}
  (\bibinfo {year} {1973})}\BibitemShut {NoStop}%
\bibitem [{\citenamefont {Kazimirov}\ \emph {et~al.}(2000)\citenamefont
  {Kazimirov}, \citenamefont {Smyk},\ and\ \citenamefont
  {Sokol’skii}}]{kazimirov2000termination}%
  \BibitemOpen
  \bibfield  {author} {\bibinfo {author} {\bibfnamefont {V.}~\bibnamefont
  {Kazimirov}}, \bibinfo {author} {\bibfnamefont {S.~Y.}\ \bibnamefont
  {Smyk}},\ and\ \bibinfo {author} {\bibfnamefont {V.}~\bibnamefont
  {Sokol’skii}},\ }\bibfield  {title} {\bibinfo {title} {Termination effect
  in x-ray diffraction studies of disordered systems},\ }\href@noop {}
  {\bibfield  {journal} {\bibinfo  {journal} {Crystallogr. Rep.}\ }\textbf
  {\bibinfo {volume} {45}},\ \bibinfo {pages} {6} (\bibinfo {year}
  {2000})}\BibitemShut {NoStop}%
\bibitem [{\citenamefont {Maley}\ \emph {et~al.}(1988)\citenamefont {Maley},
  \citenamefont {Beeman},\ and\ \citenamefont {Lannin}}]{maley1988dynamics}%
  \BibitemOpen
  \bibfield  {author} {\bibinfo {author} {\bibfnamefont {N.}~\bibnamefont
  {Maley}}, \bibinfo {author} {\bibfnamefont {D.}~\bibnamefont {Beeman}},\ and\
  \bibinfo {author} {\bibfnamefont {J.~S.}\ \bibnamefont {Lannin}},\ }\bibfield
   {title} {\bibinfo {title} {Dynamics of tetrahedral networks: Amorphous si
  and ge},\ }\href@noop {} {\bibfield  {journal} {\bibinfo  {journal} {Phys.
  Rev. B}\ }\textbf {\bibinfo {volume} {38}},\ \bibinfo {pages} {10611}
  (\bibinfo {year} {1988})}\BibitemShut {NoStop}%
\bibitem [{\citenamefont {Vink}\ \emph {et~al.}(2001)\citenamefont {Vink},
  \citenamefont {Barkema},\ and\ \citenamefont {van~der Weg}}]{vink2001raman}%
  \BibitemOpen
  \bibfield  {author} {\bibinfo {author} {\bibfnamefont {R.~L.~C.}\
  \bibnamefont {Vink}}, \bibinfo {author} {\bibfnamefont {G.~T.}\ \bibnamefont
  {Barkema}},\ and\ \bibinfo {author} {\bibfnamefont {W.~F.}\ \bibnamefont
  {van~der Weg}},\ }\bibfield  {title} {\bibinfo {title} {Raman spectra and
  structure of amorphous si},\ }\href@noop {} {\bibfield  {journal} {\bibinfo
  {journal} {Phys. Rev. B}\ }\textbf {\bibinfo {volume} {63}},\ \bibinfo
  {pages} {115210} (\bibinfo {year} {2001})}\BibitemShut {NoStop}%
\bibitem [{\citenamefont {Sellan}\ \emph {et~al.}(2010)\citenamefont {Sellan},
  \citenamefont {Landry}, \citenamefont {Turney}, \citenamefont {McGaughey},\
  and\ \citenamefont {Amon}}]{sellan2010size}%
  \BibitemOpen
  \bibfield  {author} {\bibinfo {author} {\bibfnamefont {D.~P.}\ \bibnamefont
  {Sellan}}, \bibinfo {author} {\bibfnamefont {E.~S.}\ \bibnamefont {Landry}},
  \bibinfo {author} {\bibfnamefont {J.~E.}\ \bibnamefont {Turney}}, \bibinfo
  {author} {\bibfnamefont {A.~J.~H.}\ \bibnamefont {McGaughey}},\ and\ \bibinfo
  {author} {\bibfnamefont {C.~H.}\ \bibnamefont {Amon}},\ }\bibfield  {title}
  {\bibinfo {title} {Size effects in molecular dynamics thermal conductivity
  predictions},\ }\href@noop {} {\bibfield  {journal} {\bibinfo  {journal}
  {Phys. Rev. B}\ }\textbf {\bibinfo {volume} {81}},\ \bibinfo {pages} {214305}
  (\bibinfo {year} {2010})}\BibitemShut {NoStop}%
\bibitem [{\citenamefont {He}\ \emph {et~al.}(2011{\natexlab{b}})\citenamefont
  {He}, \citenamefont {Donadio},\ and\ \citenamefont {Galli}}]{He:2011ig}%
  \BibitemOpen
  \bibfield  {author} {\bibinfo {author} {\bibfnamefont {Y.}~\bibnamefont
  {He}}, \bibinfo {author} {\bibfnamefont {D.}~\bibnamefont {Donadio}},\ and\
  \bibinfo {author} {\bibfnamefont {G.}~\bibnamefont {Galli}},\ }\bibfield
  {title} {\bibinfo {title} {{Morphology and Temperature Dependence of the
  Thermal Conductivity of Nanoporous SiGe}},\ }\href@noop {} {\bibfield
  {journal} {\bibinfo  {journal} {Nano Lett.}\ }\textbf {\bibinfo {volume}
  {11}},\ \bibinfo {pages} {3608} (\bibinfo {year}
  {2011}{\natexlab{b}})}\BibitemShut {NoStop}%
\bibitem [{\citenamefont {Clarke}\ and\ \citenamefont
  {Levi}(2003)}]{ClarkeTBC}%
  \BibitemOpen
  \bibfield  {author} {\bibinfo {author} {\bibfnamefont {D.}~\bibnamefont
  {Clarke}}\ and\ \bibinfo {author} {\bibfnamefont {C.}~\bibnamefont {Levi}},\
  }\bibfield  {title} {\bibinfo {title} {Materials design for the next
  generation thermal barrier coatings},\ }\href
  {https://doi.org/10.1146/annurev.matsci.33.011403.113718} {\bibfield
  {journal} {\bibinfo  {journal} {Annu. Rev. Mater. Res.}\ }\textbf {\bibinfo
  {volume} {33}},\ \bibinfo {pages} {383} (\bibinfo {year} {2003})}\BibitemShut
  {NoStop}%
\bibitem [{\citenamefont {Clarke}\ \emph {et~al.}(2012)\citenamefont {Clarke},
  \citenamefont {Oechsner},\ and\ \citenamefont
  {Padture}}]{clarke_oechsner_padture_2012}%
  \BibitemOpen
  \bibfield  {author} {\bibinfo {author} {\bibfnamefont {D.~R.}\ \bibnamefont
  {Clarke}}, \bibinfo {author} {\bibfnamefont {M.}~\bibnamefont {Oechsner}},\
  and\ \bibinfo {author} {\bibfnamefont {N.~P.}\ \bibnamefont {Padture}},\
  }\bibfield  {title} {\bibinfo {title} {Thermal-barrier coatings for more
  efficient gas-turbine engines},\ }\href
  {https://doi.org/10.1557/mrs.2012.232} {\bibfield  {journal} {\bibinfo
  {journal} {MRS Bulletin}\ }\textbf {\bibinfo {volume} {37}},\ \bibinfo
  {pages} {891–898} (\bibinfo {year} {2012})}\BibitemShut {NoStop}%
\bibitem [{\citenamefont {Bubnova}\ and\ \citenamefont
  {Crispin}(2012)}]{Bubnova:2012dw}%
  \BibitemOpen
  \bibfield  {author} {\bibinfo {author} {\bibfnamefont {O.}~\bibnamefont
  {Bubnova}}\ and\ \bibinfo {author} {\bibfnamefont {X.}~\bibnamefont
  {Crispin}},\ }\bibfield  {title} {\bibinfo {title} {{Towards polymer-based
  organic thermoelectric generators}},\ }\href@noop {} {\bibfield  {journal}
  {\bibinfo  {journal} {Energy Environ. Sci.}\ }\textbf {\bibinfo {volume}
  {5}},\ \bibinfo {pages} {9345} (\bibinfo {year} {2012})}\BibitemShut
  {NoStop}%
\bibitem [{\citenamefont {Ruscher}\ \emph {et~al.}(2019)\citenamefont
  {Ruscher}, \citenamefont {Rottler}, \citenamefont {Boott}, \citenamefont
  {MacLachlan},\ and\ \citenamefont {Mukherji}}]{Ruscher.3.125604}%
  \BibitemOpen
  \bibfield  {author} {\bibinfo {author} {\bibfnamefont {C.}~\bibnamefont
  {Ruscher}}, \bibinfo {author} {\bibfnamefont {J.}~\bibnamefont {Rottler}},
  \bibinfo {author} {\bibfnamefont {C.~E.}\ \bibnamefont {Boott}}, \bibinfo
  {author} {\bibfnamefont {M.~J.}\ \bibnamefont {MacLachlan}},\ and\ \bibinfo
  {author} {\bibfnamefont {D.}~\bibnamefont {Mukherji}},\ }\bibfield  {title}
  {\bibinfo {title} {Elasticity and thermal transport of commodity plastics},\
  }\href {https://doi.org/10.1103/PhysRevMaterials.3.125604} {\bibfield
  {journal} {\bibinfo  {journal} {Phys. Rev. Materials}\ }\textbf {\bibinfo
  {volume} {3}},\ \bibinfo {pages} {125604} (\bibinfo {year}
  {2019})}\BibitemShut {NoStop}%
\bibitem [{\citenamefont {Cappai}\ \emph {et~al.}(2020)\citenamefont {Cappai},
  \citenamefont {Antidormi}, \citenamefont {Bosin}, \citenamefont {Narducci},
  \citenamefont {Colombo},\ and\ \citenamefont {Melis}}]{Cappai:2020bh}%
  \BibitemOpen
  \bibfield  {author} {\bibinfo {author} {\bibfnamefont {A.}~\bibnamefont
  {Cappai}}, \bibinfo {author} {\bibfnamefont {A.}~\bibnamefont {Antidormi}},
  \bibinfo {author} {\bibfnamefont {A.}~\bibnamefont {Bosin}}, \bibinfo
  {author} {\bibfnamefont {D.}~\bibnamefont {Narducci}}, \bibinfo {author}
  {\bibfnamefont {L.}~\bibnamefont {Colombo}},\ and\ \bibinfo {author}
  {\bibfnamefont {C.}~\bibnamefont {Melis}},\ }\bibfield  {title} {\bibinfo
  {title} {{Impact of synthetic conditions on the anisotropic thermal
  conductivity of poly(3,4-ethylenedioxythiophene) (PEDOT): A molecular
  dynamics investigation}},\ }\href@noop {} {\bibfield  {journal} {\bibinfo
  {journal} {Phys. Rev. Materials}\ }\textbf {\bibinfo {volume} {4}},\ \bibinfo
  {pages} {035401} (\bibinfo {year} {2020})}\BibitemShut {NoStop}%
\end{thebibliography}
%

\end{document}